\documentclass[letterpaper, 10 pt, conference]{IEEEtran}  

\IEEEoverridecommandlockouts                              

\usepackage{graphics} 
\usepackage{epsfig} 
\usepackage{mathptmx} 
\usepackage{times} 
\usepackage{amsmath} 
\usepackage{authblk}
\usepackage{outlines}
\usepackage[dvipsnames]{xcolor}
\usepackage{cite}
\usepackage{graphicx}
\usepackage[T1]{fontenc}
\usepackage[utf8]{inputenc} 
\usepackage{url}

\ifCLASSOPTIONcompsoc
    \usepackage[caption=false, font=normalsize, labelfont=sf, textfont=sf]{subfig}
\else
\usepackage[caption=false, font=footnotesize]{subfig}
\fi
\newsubfloat{figure}
\usepackage{etoolbox}
\makeatletter
\patchcmd{\@makecaption}
  {\scshape}
  {}
  {}
  {}
\makeatother

\title{\LARGE \bf
Hit-time and hit-position reconstruction in strips of plastic scintillators using multi-threshold readouts
}
\author{N.~G.~Sharma$^{1}$, M.~Silarski$^1$,
J.~Chhokar$^1$, E.~Czerwi\'nski$^1$, C.~Curceanu$^{2}$, K.~Dulski$^1$,  K.~Farbaniec$^1$, A.~Gajos$^1$, \\ R.~Del ~Grande$^2$, 
M.~Gorgol$^3$, B.~C.~Hiesmayr$^4$, B.~Jasi\'nska$^3$, K.~Kacprzak$^1$, \L.~Kap\l on$^1$, D.~Kisielewska$^1$, \\ 
K.~Klimaszewski$^6$, G.~Korcyl$^1$, 
 P.~Kowalski$^6$, N.~Krawczyk$^1$, W.~Krzemie\'n$^5$, T.~Kozik$^1$, E.~Kubicz$^1$, M. Mohammed$^{1,7}$, \\
 Sz.~Nied\'zwiecki$^1$, 
 M.~Pa\l ka$^1$, M.~Pawlik-Nied{\'z}wiecka$^1$, L.~Raczy\'nski$^6$, J.~Raj$^1$, S.~Sharma$^1$, \\ S.~Shivani$^1$,
R.~Y.~Shopa$^6$, M.~Skurzok$^1$, W.~Wi\'slicki$^6$, B.~Zgardzi{\'n}ska$^3$, P.~Moskal$^1$ \\
\thanks{$^1$ Faculty of Physics, Astronomy and Applied Computer Science, Jagiellonian University, 30-348 Cracow }
\thanks{$^2$INFN, Laboratori Nazionali di Frascati, 00044 Frascati, Italy}
\thanks{$^3$ Institute of Physics, Maria Curie-Sk\l odowska University, 20-031 Lublin, Poland}
\thanks{$^4$ Faculty of Physics, University of Vienna, 1090 Vienna, Austria}
\thanks{$^5$ High Energy Physics Division, National Centre for Nuclear Research, 05-400 Otwock-\'Swierk, Poland}
\thanks{$^6$ Department of Complex Systems, National Centre for Nuclear Research, 05-400 Otwock-\'Swierk, Poland}
\thanks{$^7$ Department of Physics, College of Education for Pure Sciences, University of Mosul, Mosul, Iraq
}
}

\begin{document}
\bstctlcite{IEEEexample:BSTcontrol}
\maketitle
\thispagestyle{empty}
\pagestyle{empty}

\begin{abstract}

In this article a new method for the reconstruction of hit-position and hit-time of photons in long scintillator detectors is investigated. This research is motivated by the recent development of the positron emission tomography scanners based on plastic scintillators. The proposed method constitutes a new way of signal processing in Multi-Voltage-Technique. It is based on the determination of the degree of similarity between the registered signals and the synchronized model signals stored in a library. The library was established for a set of well defined hit-positions along the length of the scintillator. The Mahalanobis distance was used as a measure of similarity between the two compared signals. The method was validated on the experimental data measured using two-strips J-PET prototype with dimensions of 5x9x300 mm$^3$. The obtained Time-of-Flight (TOF) and spatial resolutions amount to 325~ps (FWHM) and 25~mm (FWHM), respectively. The TOF resolution was also compared to the results of an analogous study done using Linear Fitting method. The best TOF resolution was obtained with this method at four pre-defined threshold levels which was comparable to the resolution achieved from the Mahalanobis distance at two pre-defined threshold levels. Although the algorithm of Linear Fitting method is much simpler to apply than the Mahalanobis method, the application of the Mahalanobis distance requires a lower number of applied threshold levels and, hence, decreases the costs of electronics used in PET scanner.
\end{abstract}

\section{Introduction}

Positron Emission Tomography is a noninvasive imaging technique used in medical diagnostics. It uses a short-lived positron-emitting radioactive tracer which is injected into the patient's body. The density distribution of the tracer is determined by pairs of back-to-back gamma quanta registered by detectors surrounding the patient. Emission of gamma quanta occurs as a result of annihilation of a positron with an electron present inside the patient's body. PET scanners available commercially use crystal scintillators as radiation detectors and they have an axial extent of about 17-25~cm~\cite{VAN2016,SLOMKA2016}. The time and spatial resolutions obtained by the best PET scanners are about 210-400~ps and 4.7-6~mm, respectively~\cite{SLUIS2019,VAN2016}.

One of the challenges in the PET tomography is to increase the axial extent. In current PET scanners roughly 85$\%$-90$\%$ of the patient's body is outside of the field-of-view~(FOV) of the scanner, hence, only 1$\%$ of pairs of coincidence photons emitted from the body are detected~\cite{VAN2016,SLOMKA2016,SZYMON2017}. The extension of detector rings from 20~cm axial FOV to a 200~cm FOV would allow maximal detection of radiation emitted from the body and hence improve the sensitivity and signal-to-noise ratio. Furthermore, with improved sensitivity the radiation dose needed for the whole body scan could be decreased which in turn will allow usage of short-lived radionuclides like $^{11}${}C. The extension of axial FOV to total-body scan may also enable newly developed positronium imaging~\cite{MOSKAL2019A,MOSKAL2019B}. To address this problem many new PET concepts have been proposed, like the Lead-Walled Straw PET detector (LWS)~\cite{LWS2001,LWS2005}, the Resistive Plate Chamber PET (RPC)~\cite{RPC2006,RPC2009}, axial geometry based PET scanners~\cite{CIMA2006,AXPET2014} and the EXPLORER total-body PET tomograph~\cite{CHERRY2017,VISHWANATH2017,CHERRY2017A}.
The J-PET scanner~\cite{Moskal2011A,MOSKAL14C,Moskal2015A,Moskal2014A,SYMRSKI2017} constitutes another economical solution with large FOV, built out of plastic scintillators. The relatively low probability of annihilation photons detection with plastic scintillators can be overcome by using longer modules and more detection layers~\cite{MOSKAL2016A}. The axial arrangement of plastic strips allows placing the electronic readout system~\cite{MAREK2014A,MAREK17A,GREG2016A,GREG2018A} outside of the detectors, with a great benefit for medical diagnostics since it enables simultaneous application of PET and CT as well as PET and MRI. The later hybrid is already achieved also with the crystal based PET/MRI systems~\cite{VAN2016}.

 
This novel solution of long detection modules requires new optimization methods for the determination of the hit-time and the hit-position. There have been several attempts made so far to optimize the reconstruction~\cite{Moskal2015A,Lech2014B,Lech2015A,NEHA2015A,NEHA2015B,LECH2017A} however, in this article we present another idea based on the fact that the shape of signals registered at the end of a plastic bar depends significantly on the position of interaction. The idea was realized in two steps: first, creation of library of synchronized model signals and second, reconstruction of hit-time and hit-position of annihilation gamma quanta. The reconstruction is based on the comparison between measured signal and signals from the library. As a measure of similarity a Mahalanobis distance is used which accounts for the uncertainties of the signals samplings and correlations between uncertainties. A similar method was already presented in~\cite{NEHA2015A} but with chi-square minimization function used as a measure of similarity. In the present article the first step is extended further with additional conditions of signal selection followed by the reconstruction method using Mahalanobis distance. The method, similarly as the one introduced in~\cite{NEHA2015A}, requires application of the Multi-Voltage Threshold (MVT) technique introduced and developed by~\cite{XIE2005,XIE2015,XIE2016}. The main motivation of the research presented in this article is to find out a measure of similarity which would enable to minimize the number of threshold used, and to facilitate more economical construction of PET modalities with long axial FOV.\\  
The article is structured into four sections: First, the concept of the J-PET detector and the two strip J-PET prototype is briefly described. Then, the general idea behind the proposed reconstruction method including a description of signal processing is presented. In the next section, the realization  and validation of the method is explained. Finally, the experimental results are presented and discussed.

\section{The J-PET detector}
\begin{figure}[thpb]
      \centering
      \includegraphics[scale=0.30]{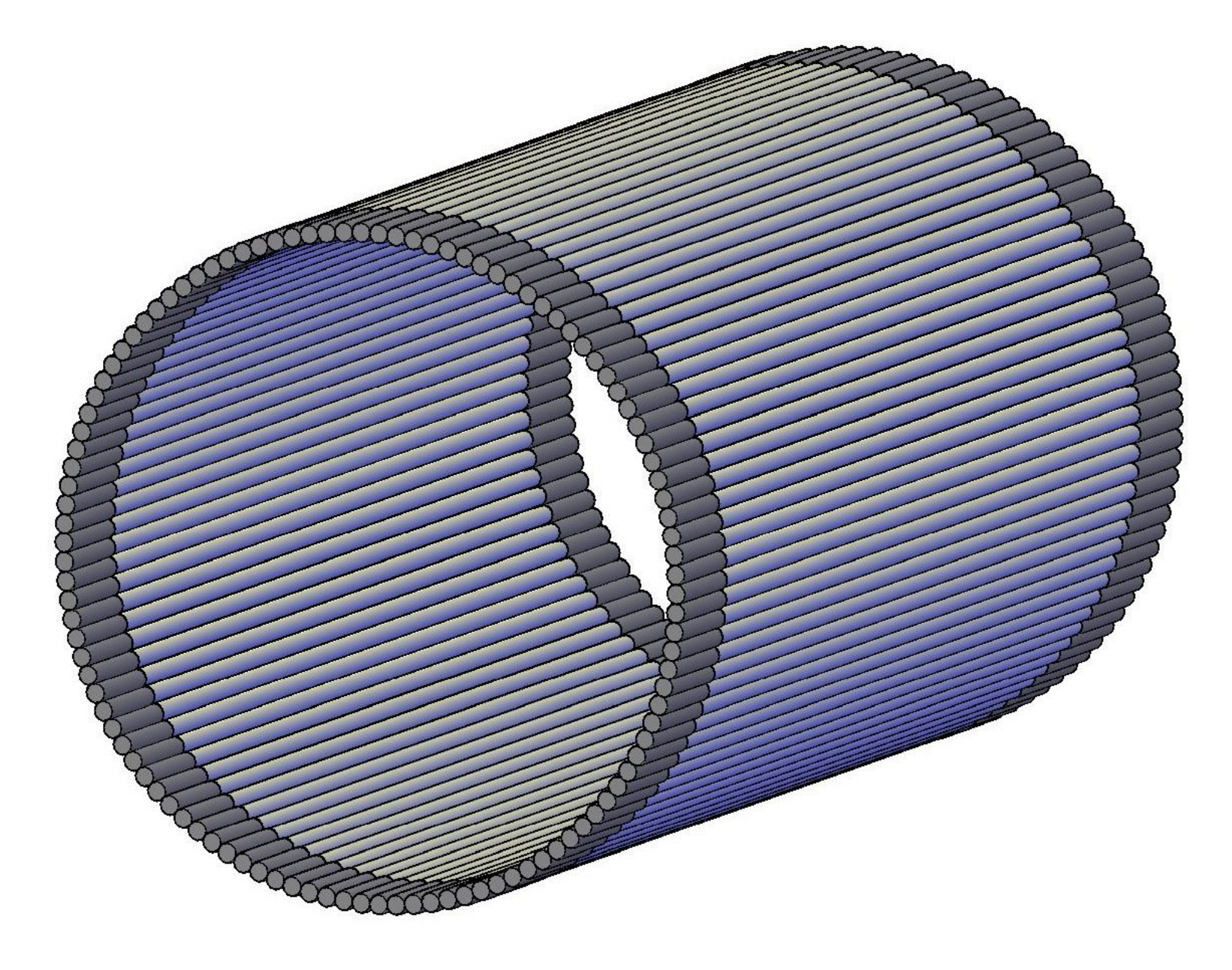}
      \caption{Schematic arrangement of the detection modules providing large field-of-view (FOV) used in the J-PET scanner. In the module all the scintillators are connected to a pair of photomultipliers at both ends.}
      \label{fig:Fig1}
\end{figure}

\begin{figure}[thpb]
      \centering
      \includegraphics[scale=0.31]{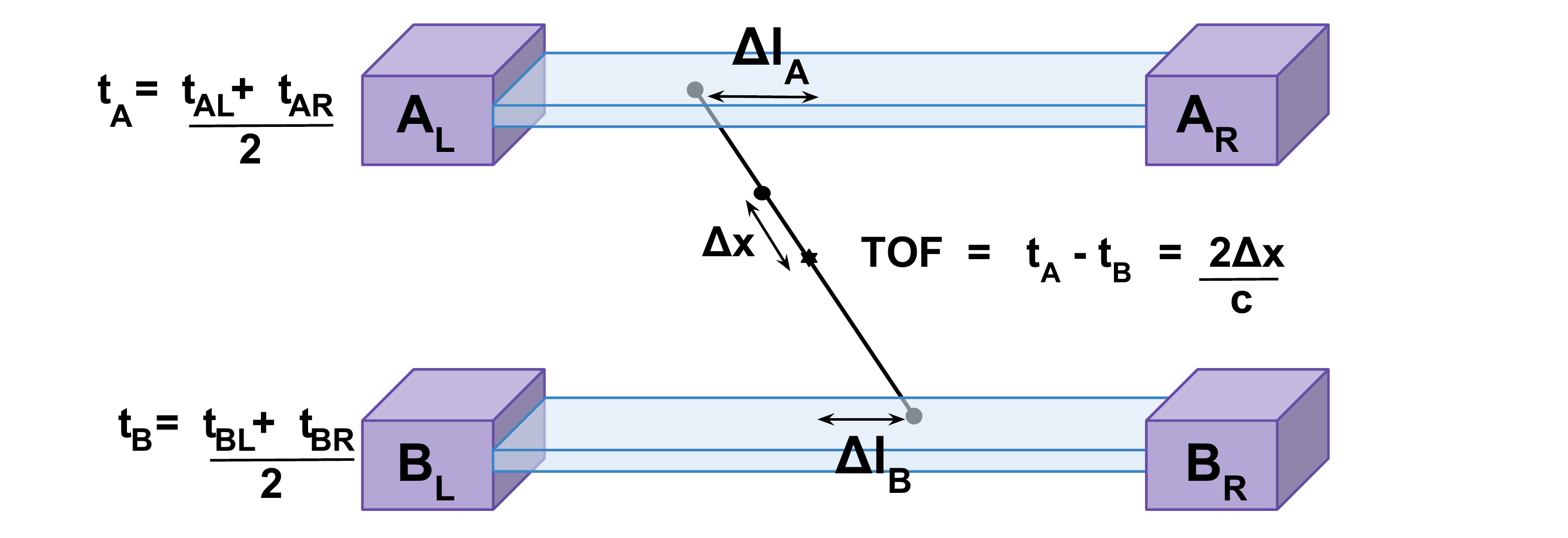}
      \caption{Pictorial representation of the annihilation point reconstruction using the Time-of-Flight (TOF) method in the J-PET detector. t$_{AL}$ and t$_{AR}$ are the time values measured for signals obtained from left- and right-ended photomultipliers of strip A at some pre-defined threshold levels and t$_A$ is their average. Similarly, t$_{BL}$ and t$_{BR}$ are the time values of left and right signals obtained from strip B and t$_B$ is the average. $\Delta$ l$_A$ and $\Delta$ l$_B$ are the distances of the hit-positions of photons within the scintillators A and B from their central positions, respectively. $\Delta$ $x$ is the distance along LOR between the point of annihilation and the center of LOR.}
     \label{fig:Fig2}
\end{figure}

The J-PET scanner uses organic scintillators arranged axially (see Fig.\ref{fig:Fig1}) as a radiation detectors. Signals from each detection module are read out by a pair of photomultipliers connected to the two ends of each strip, as shown in Fig.\ref{fig:Fig2}. Organic scintillators have a long light attenuation length (in the order of 100~cm to 400~cm~\cite{SCINOX,BATTALECTURE}), in comparison to crystal scintillators ($\sim$ 10~cm~\cite{BATTALECTURE,SAINTGOBAIN,MAO2008}), which allows to make large diagnostic chambers from long scintillators strips. The annihilation gamma quanta with 511~keV energy interact in plastic scintillators predominantly via Compton scattering. The hit-position and hit-time of the annihilated photons can be determined from the arrival time information of signals detected by the photomultipliers placed at each end of the detection module. Similarly, the position along the line-of-response (LOR) can be determined using Time-of-Flight information, as shown in Fig.~\ref{fig:Fig2}. 
   
\begin{figure}[h]
    \centering    
    \includegraphics[scale=0.35]{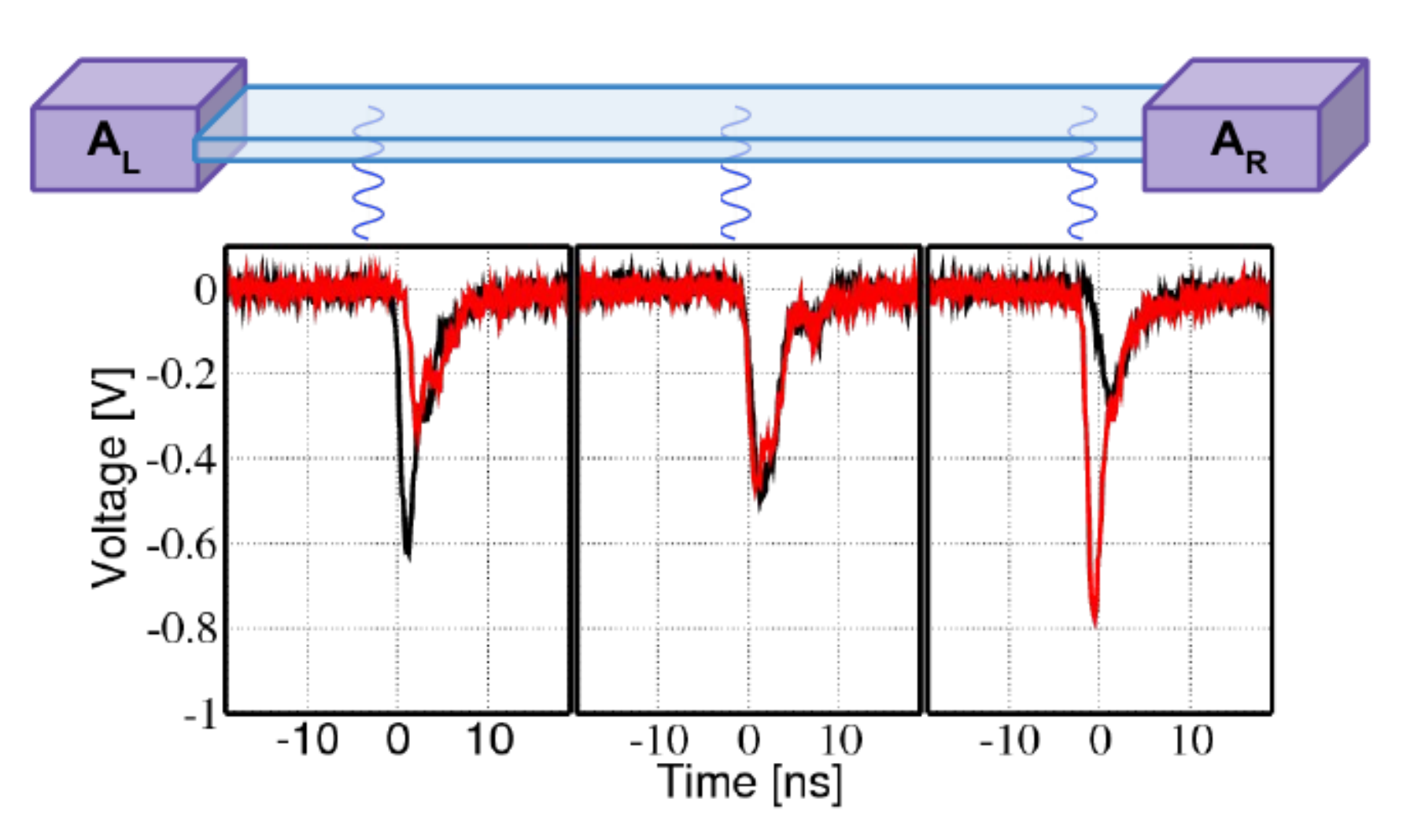}
    \caption{ Example of signals obtained at three different irradiated positions from a pair of photomultipliers connected to the ends of a scintillator strip. Solid black and red lines represent the left and right signals measured by left (A$_L$) and right (A$_R$) photomultiplier, respectively. The measurements were conducted with BC-420 scintillators~\cite{SAINTGOBAIN} and R9800
    photomultipiers~\cite{HAMAMATSU}.}
    \label{fig:Fig3}
\end{figure}

\begin{figure}
    \centering    
    \includegraphics[scale=0.35]{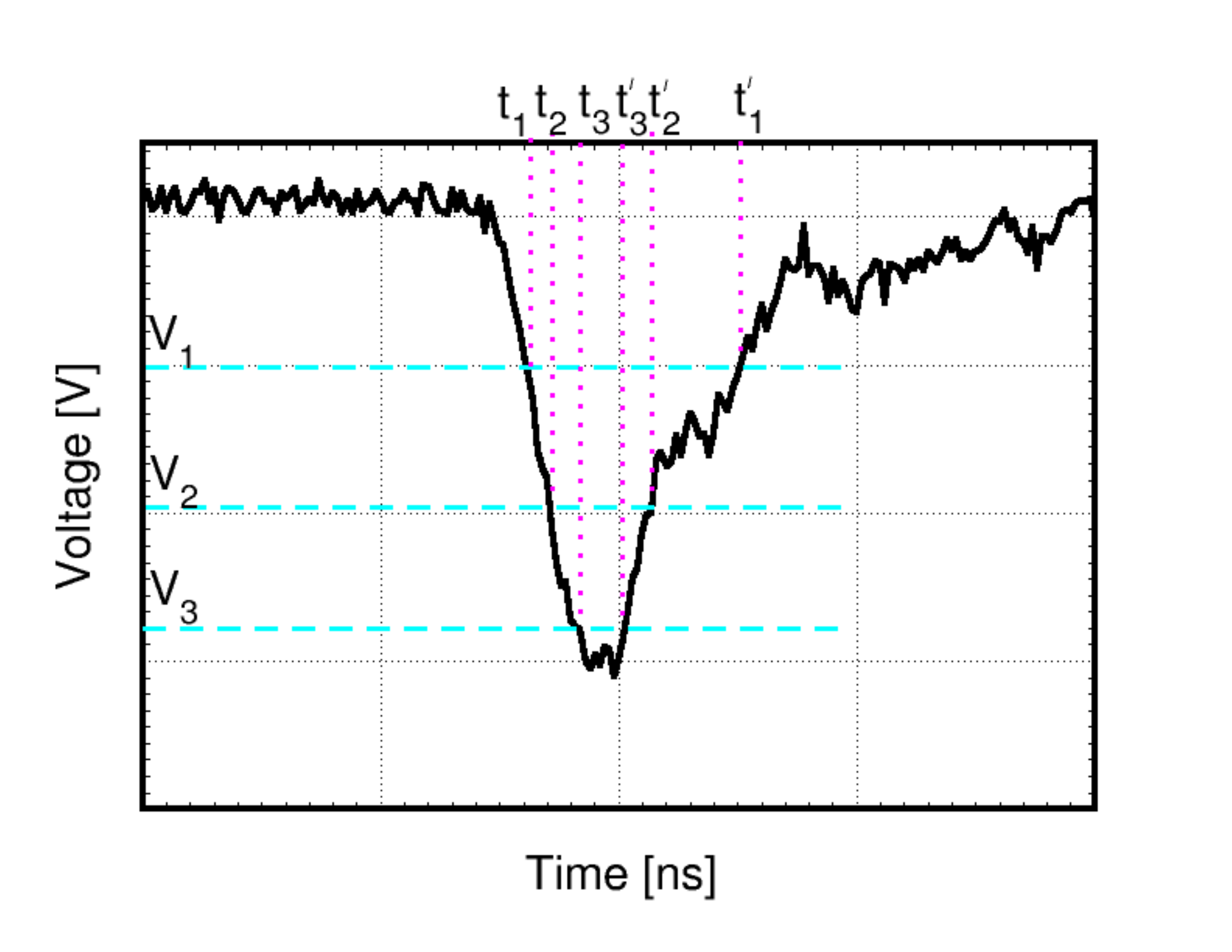}
    \caption{Sampling of signal in the voltage domain at three defined voltages V$_1$, V$_2$ and V$_3$ as it is done by the J-PET electronics. As a result, in the presented example one obtains three time values measured at the leading
    (t$_{1}$, t$_{2}$, t$_{3}$) and three times values at the trailing edge (t$^{'}_{1}$, t$^{'}_{2}$, t$^{'}_{3}$).}
    \label{fig:Fig4}
\end{figure}

Fig.~\ref{fig:Fig3} shows typical signals obtained from a single detection module consisting of a plastic scintillator optically connected to a pair of photomultipliers at three different irradiated positions with gamma quanta. As one can see the shape and amplitude of the signals vary with the interaction point of gamma quanta within the scintillator. 
The J-PET electronics used to process signals from the photomultipliers enables to determine their widths and times at which they crossed the given reference voltages by means of multi-threshold constant-level discriminators~\cite{MAREK2014A,MAREK17A}. A pictorial representation of signal sampling in voltage domain is shown in Fig.\ref{fig:Fig4}.

As mentioned above and illustrated in Fig.\ref{fig:Fig3}, in the J-PET scanner the shape and amplitude of signals strongly depend on the hit-position of gamma quanta in the strip. This dependency becomes stronger with increasing size of the scintillator and hence, influences the time and position resolution. Moreover, the final uncertainty in reconstruction of the annihilation point and herewith the performance of the PET-scanner depends strongly on the time resolution.\\
As it will be shown in next sections of this article, the strong change of signal shape along the scintillator strip can be utilized in optimizing the position and time of photons interaction. It is important to remark that a single event as a result of annihilation, refers to two signal curves in each strip, so in total of four signal curves. The reconstruction method which we propose was developed and optimized on the two strips J-PET prototype read out by a Serial Data Analyzer which provided full signals gathering. However, in order to simulate the real J-PET tomograph electronics, we have emulated the sampling in the voltage domain~\cite{NehaPhdThesis}. In general this reconstruction method can be used in other state of the art scanners, which sample signals in voltage domain either by means of multi-threshold constant-level discriminators or by means of constant-fraction discriminators. Here it is worth noting that recently new methods for multi-voltage sampling based on Field Programmable Gate Arrays (FPGA) were developed~\cite{XI2013,XIE2013A,MAREK2014A,MAREK17A,GREG2016A,GREG2018A,JUN2018,JUN2016A,JUN2016B} for application in Positron Emission Tomography. This type of system exploits FPGAs differential inputs as comparators. In contrast to standard approach~\cite{KIM2009B}, where external comparator chips are used, it allows to create more compact and less expensive systems. The process of analog signal comparison in the differential buffer is  similar to the usage of standard comparator. On the positive input of an FPGA differential buffer a predefined voltage from data acquisition system (DAQ) is delivered and on the negative side a measured signal. When the signal is crossing the threshold level given by the DAC it creates a transition of digital signal inside the FPGA. This transition corresponds to either a leading or trailing edge and precise times, corresponding to the arrival of these edges, are measured with time to digital Converters (TDC)~\cite{KALISZ1997A,WU2008A} implemented within FPGA device. It has been shown in~\cite{NEISER2013} that with this approach the intrinsic TDC precision of 20 ps rms per channel is achievable.

\section{Hit-position and hit-time reconstruction method for long plastic scintillators}
\subsection{Realization of the method}
The proposed reconstruction method, as mentioned earlier, utilizes the signal shape to determine the hit-time and hit-position of photons registered in long scintillating detectors. To this end the detector is first characterized by determination of signal shapes at a set of well-defined hit positions along the scintillator. Since the signal amplitude can vary even at the same hit position (e.g. due to different energy deposition) we synchronize and average a big statistics of signals at each hit-position which defines so called model events. Having such a standardized detector the reconstruction is done by comparing the registered event to all the model events stored in a database.  
\begin{figure}
  \centering
  \includegraphics[scale=0.3]{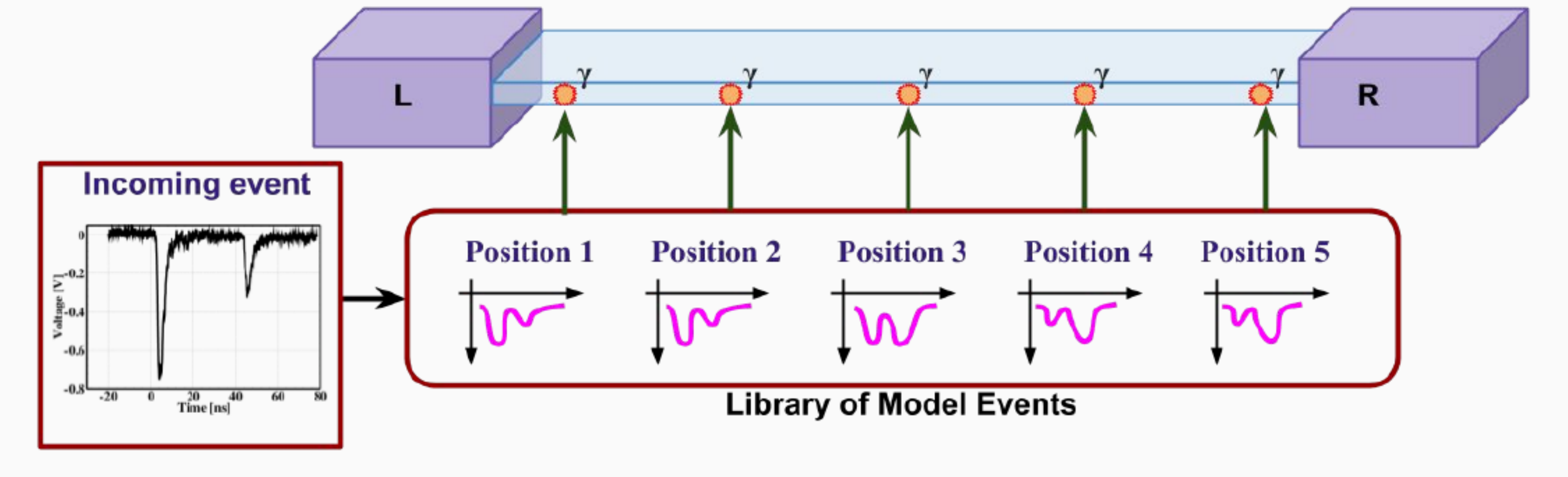}
  \caption{A schematic representation of the developed reconstruction method. The measured event, consisting of signals registered at both ends of the scintillator is compared to model events determined for a set of well defined positions along the detector. Orange dots are representing the points of interactions of gamma quanta within the scintillator.}
  \label{fig:Fig5}
  \end{figure}
The known hit-time and hit-position of the most similar model event in the database is then treated as the hit-time and hit-position of registered event. In Fig.\ref{fig:Fig5} a schematic illustration of the working principle behind the presented reconstruction method is shown. The Mahalanobis distance formula~\cite{MAHA1936} is used as a measure of similarity between the two compared events. It is a measure of the deviations of the different mean values in terms of the standard deviation in a multivariate analysis. Here, deviation is defined as the correlation between the model and registered events~\cite{NehaPhdThesis}.

\subsection{Library of synchronized model events}\label{library}

The library was determined from a scan of scintillator strip with a collimated beam of annihilation photons with profile width of $\sim$1.5 mm~\cite{EWELINA2016A}. The strips were scanned with a step of 3~mm, such that there were in total 98 scanned positions and the information of the irradiated position was obtained by the synchronization of collimator movement with the data acquisition system~\cite{GREG2016A}. High statistics of events ($\sim$5000) were collected for each irradiated position. The construction of model signals library was already described in details in~\cite{NEHA2015A}, here we present only main steps of the procedure comprising:\\
\begin{itemize}
    \item  {Synchronization of signals: For each hit-position of photons all the collected events were synchronized in order to have the same hit-time value. This was obtained by using a calibration constant, t$_{synch}$ = (t$_L$ + t$_R$)/2 for each event, where t$_L$ and t$_R$ are the time values of the left and right ended signals calculated at their beginning point, respectively, and the signals were shifted in time by t$_{synch}$ such that the resulted synchronized signal time is equal to zero as shown in Fig.\ref{fig:Fig6} at three different irradiated positions.}
\begin{figure}
   \vspace{-0.5mm}
    \centering    
    \includegraphics[scale=0.33]{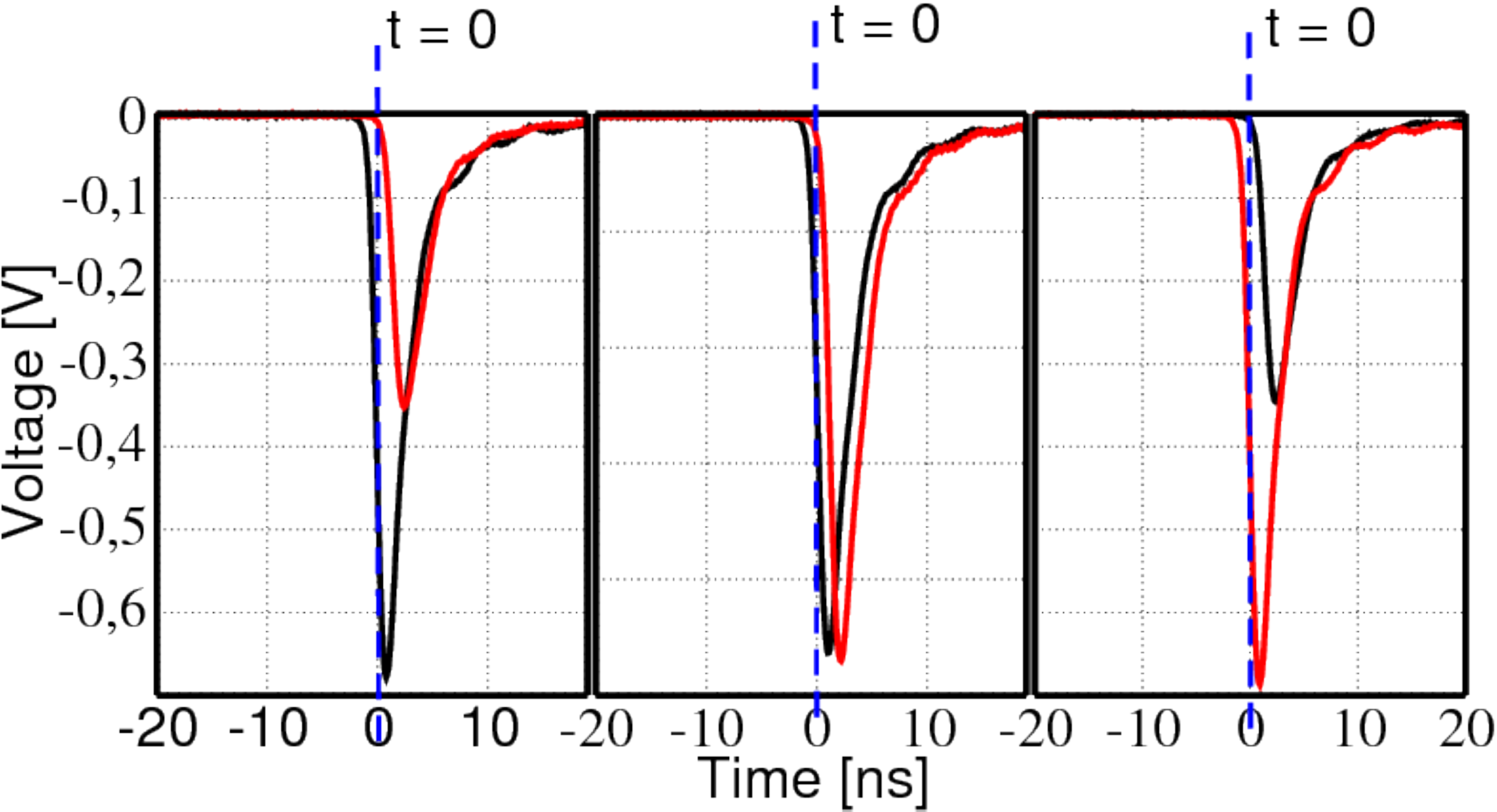}
    \caption{ Example of synchronized signals at three different hit-positions. Solid black and red lines represent signals measured by left and right photomultiplier, respectively.}
    \label{fig:Fig6}
\end{figure}

\item {Determination of average event: Then averaging over all events for each hit-position was performed and the obtained average event was treated as a reference to align the measured signal as it is shown in Fig.~\ref{Fig7a} and Fig.~\ref{Fig7b}}.

\item {Event's alignment: The $\chi^2$ statistics was used for the alignment of signals which is necessary to suppress the spread of the events in terms of amplitude and time:
\begin{equation}\label{eq:Equation1}
\begin{aligned}
\chi ^2 (\delta t,\alpha _{L,} \alpha _R ) = & \sum\limits_{i = 1}^n
{\frac{{(t_{AvgLeft} (V_i ) - t_{dbLeft} (\alpha _L V_i ) - \delta
t)}}{n}} ^2 + \\
& \sum\limits_{i = 1}^m {\frac{{(t_{AvgRight} (V_i )
- t_{dbRight} (\alpha _R V_i ) - \delta t)}}{m}} ^2
\end{aligned}
\end{equation}
Here, $\delta$t is the shift along the time axis and $\alpha_L$, $\alpha_R$ are the normalization factors for signals registered at both ends of the scintillator (left and right, respectively) as shown in Fig.~\ref{Fig7a}. $t_{AvgLeft}(V_i)$ and $t_{AvgRight}(V_i)$ denote the time of left and right average signals computed for voltage $V_i$ at their leading edge. $t_{dbLeft}(\alpha_{L}V_i)$ and $t_{dbRight}(\alpha_{R}V_i)$ are the times computed for rescaled left and right signals at their leading edge, respectively~\cite{NEHA2015A,NEHA2015B}. Finally, \textit{n} and \textit{m} denote the number of points sampled at the leading edge of the left and right signals, respectively.}
\begin{figure}[h]
\centering
  \subfloat[\label{Fig7a}]{\includegraphics[scale=0.33]{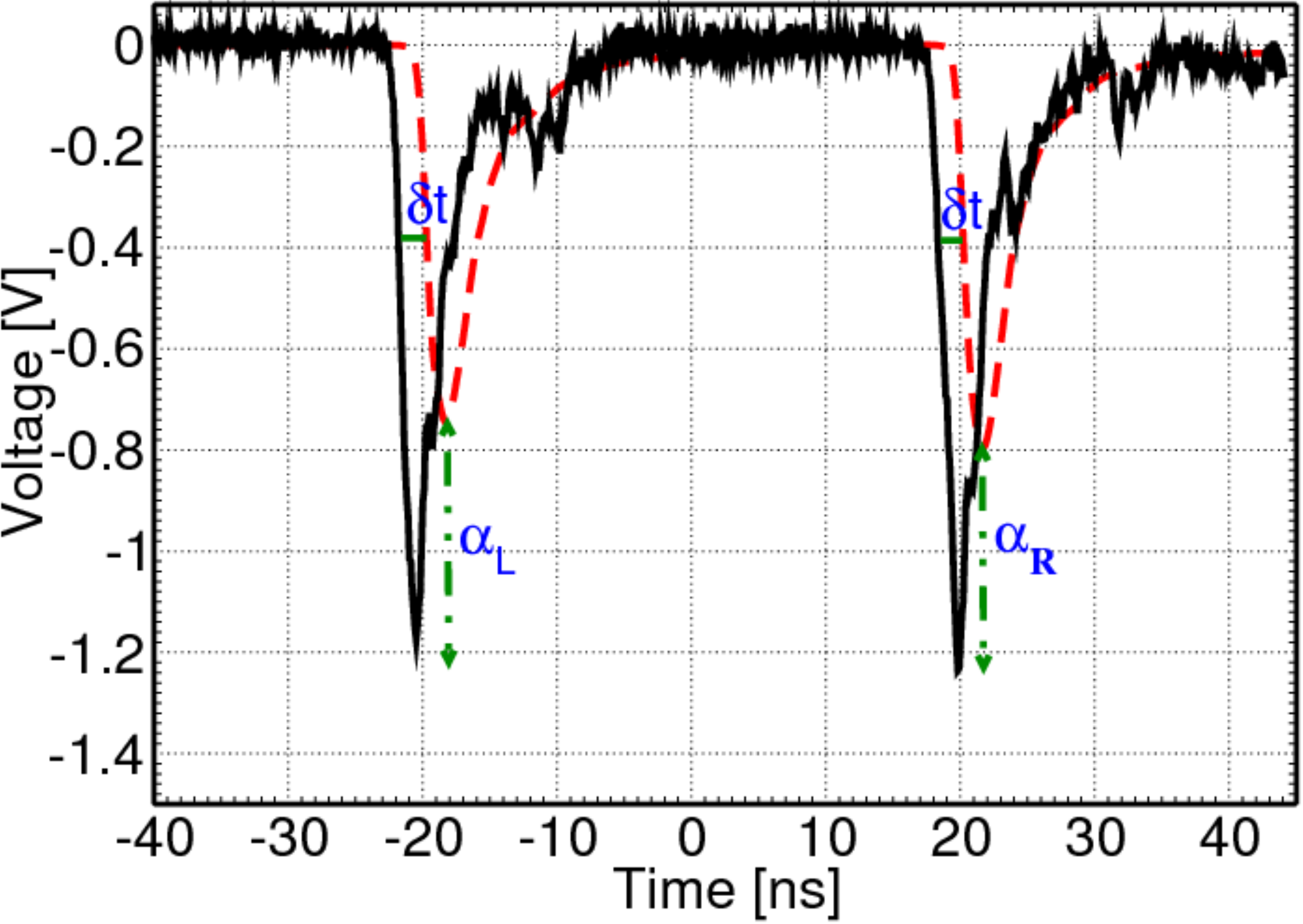}}
  \hfill
  \subfloat[\label{Fig7b}]{\includegraphics[scale=0.33]{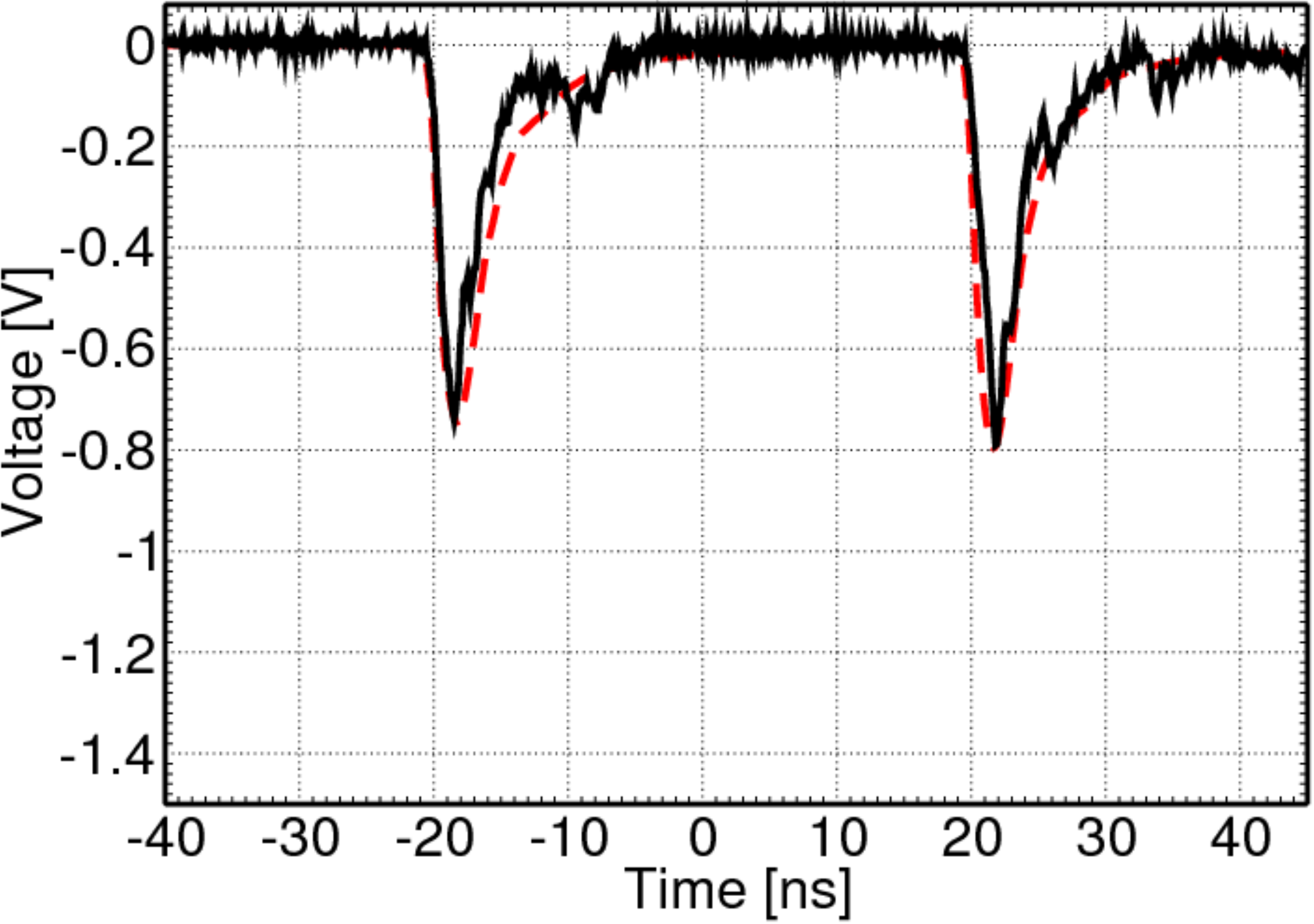}}
    \caption{(a) Example of database event before the alignment to average signals.  (b) The same database event after the alignment to average signals using $\alpha_L$, $\alpha_R$ and $\delta$t parameters. Black curve represents measured events while the red one represents computed average events. $\delta$t is the shift along the time axis and $\alpha_L$, $\alpha_R$ are the normalization factors for signals registered at both ends of the scintillator (left and right, respectively. }
    \label{fig:Fig7}
\end{figure}
{The set of parameters $\alpha_L$, $\alpha_R$ and $\delta$t was used for alignment of the measured events for which $\chi^2$ was minimal.}
\item {Determination of model events shape: Afterwards an averaging of all the rescaled events was performed resulting in determination of the so called model event. In Fig.\ref{fig:Fig9} exemplary model events obtained for three different hit-positions are shown.}
\begin{figure}[h]
    \centering    
    \includegraphics[scale=0.33]{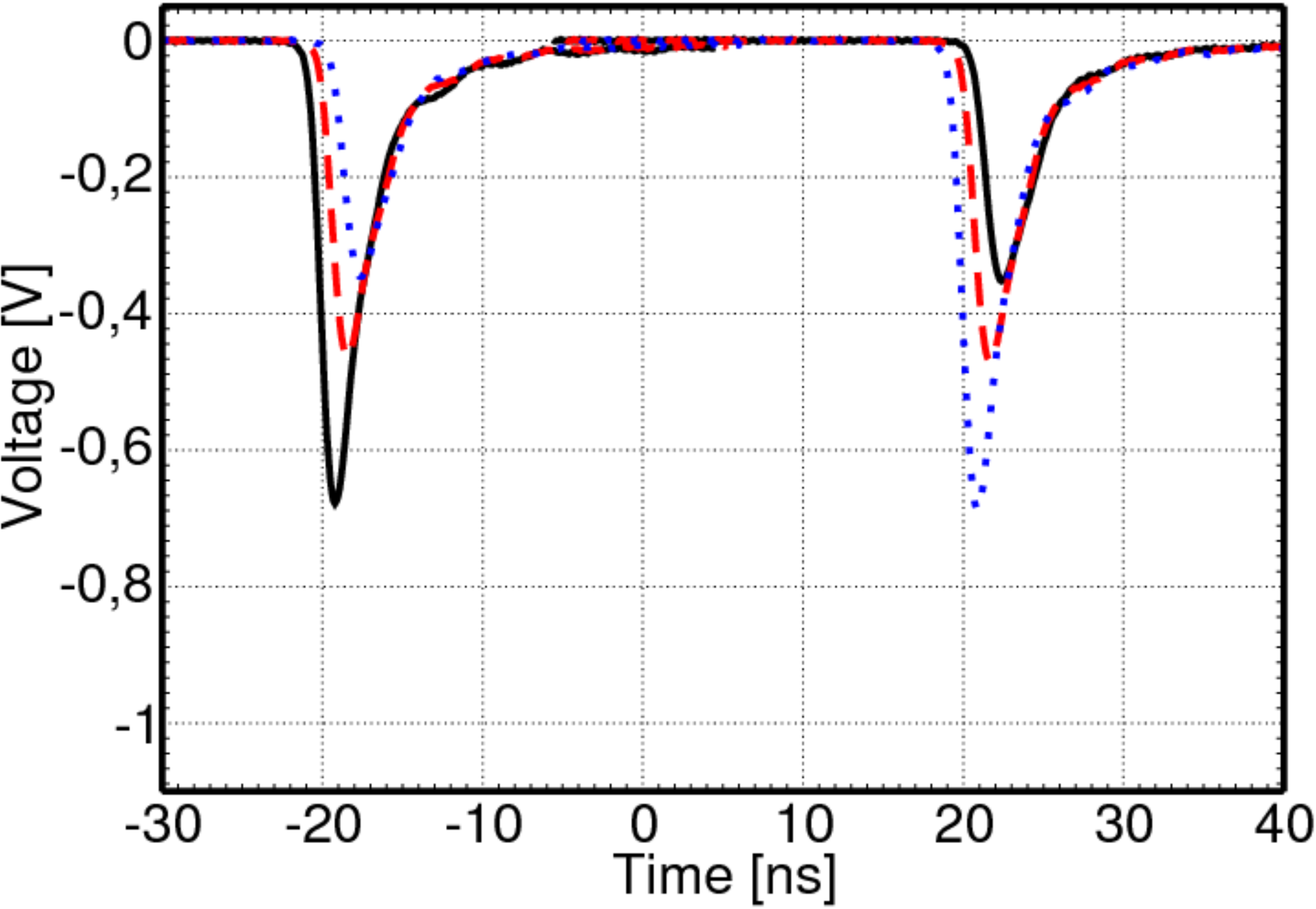}
    \caption{Exemplary model signals at three different irradiated positions along a strip. Black solid curve shows the model event produced
for the position nearest to the left end of the strip. Red dashed curve denotes the model event belonging to the central hit-position and blue dotted curve is the model event for the position lying in the proximity of the right end of the scintillator strip.}
    \label{fig:Fig9}
\end{figure}
\end{itemize}

\subsection{Reconstruction of hit-position and hit-time}
The hit-position of gamma quanta is reconstructed exploiting the Mahalanobis distance~\cite{MAHA1936} as 
a measure of similarity for two compared events (the measured and the model one), using Eq.~\ref{eq:Equation2}:
\begin{equation}
\label{eq:Equation2}
	M.D(z,\Delta t) = \sqrt{\vec{x}(z, \Delta t) C(z)^{ - 1}\vec{x}(z, \Delta t)^{T}}~
\end{equation}
where \textit{z} is the hit-position along the length of the scintillator and \textit{$\Delta$t} denotes the time shift between the two compared events. $\vec{x}(z, \Delta t ) $ is a vector whose elements equal to the differences between the elements of the measured and the model event vectors shifted by \textit{$\Delta t$} and \textit{C(z)} are the covariance matrices. As a result of comparison, only those values of \textit{z} and \textit{$\Delta$t} were taken into consideration for which minimum value of the Mahalanobis distance was obtained. The minimum value corresponds to the most similar model event. Formulation of covariance matrices \textit{C(z)} and $\vec{x}(z, \Delta t ) $ vectors was as follows:
\begin{itemize}
    \item {x-vector: The elements of the measured and the model event vectors are the time values calculated at defined set of threshold levels for signals registered at both detector sides. The number of elements in $\vec{x}$ depends on the number of threshold levels applied to the signals of an event. For example, $\vec{x}(z, \Delta t ) $ of an event at a single applied threshold level will consist of two elements: one measured from the left- and right-ended photomultipliers connected to the scintillator. It is defined as:
\begin{equation}
\label{eq:x-vector}
    \vec{x}(z, \Delta t ) =  [\vec{x^L}(z, \Delta t), \vec{x^R}(z, \Delta t)]
\end{equation}
 and
\[
\vec{x^L}(z, \Delta t)  = t^{L}  - t^{Lmod}(z) - \Delta t
\]
\[
\vec{x^R}(z, \Delta t)  = t^{R}  - t^{Rmod}(z) - \Delta t
\]
where \textit{t$^{L}$} and \textit{t$^{R}$} are the elements of $\vec{t}$ whose values are times at the applied threshold for the left and right signal of measured event, respectively. Similarly, \textit{t$^{Lmod}$} and \textit{t$^{Rmod}$} are the elements of $\vec{t^{mod}}$ whose values are the times obtained at the same threshold for model event:
\[
\vec t  = \left[ {t^{L}} \right. , \left. {t^{R}} \right]
\]
\[
\vec t^{mod}  = \left[ {t^{Lmod}} ,{t^{Lmod}} \right]~
\]
\item Covariance matrix : The covariance matrix C$(z)$, was computed for each scanned position (mentioned in sec.\ref{library}) for the defined set of threshold levels using Eq.\ref{eq:Equation4};
\begin{equation}\label{eq:Equation4}
C_{ij}  = \sum\limits_{k = 1}^N {\frac{{(\vec{t}_{k(i)} -
\vec{t}_{avg(i)} )(\vec{t}_{k(j)} - \vec{t}_{avg(j)} )}}{N}}
\end{equation}
where $\vec{t_k}$ is the vector of times for the $k^{th}$ measured event belonging to the defined hit-position from the scan and $\vec{t}_{avg}$ is the average times vector determined for the same hit-position. Indices \textit{i} and \textit{j} enumerate the applied threshold levels and \textit{N} denotes number of events corresponding to the position in the database. The number of elements of the covariance matrix is equal to $(2m)^2$, where $m$ denotes the number of threshold levels applied to the signals.
\item Hit-time: Hit-time (interaction time of gamma quanta) is equal to $\Delta$t, time shift between the two compared events~\cite{MOSKALPATENT} for minimum Mahalanobis distance obtained from Eq.\ref{eq:Equation2}. Thus, Time-of-Flight of annihilation gamma quanta registered in a pair of detectors, denoted A and B, is equal to the difference between hit-times in these detectors and is computed by:
\begin{equation}
\label{eq:Equation5}
TOF = \Delta t_A - \Delta t_B
\end{equation}
It is worth to mention that the obtained TOF is independent of the trigger time because the same time affects both detectors. Hence, the proposed method allows direct determination of line-of-response (LOR) and Time-of-Flight.
\item Hit-position: Hit-position is the position \textit{(z)} of most similar model
event from the library with respect to measured event for which minimal value of Mahalanobis distance is obtained from Eq.\ref{eq:Equation2}. This value of \textit{(z)} is called as the reconstructed hit-position.
}
\end{itemize}

\section{Linear-Fit method}
\begin{figure}[h]
    \centering    
    \includegraphics[scale=0.33]{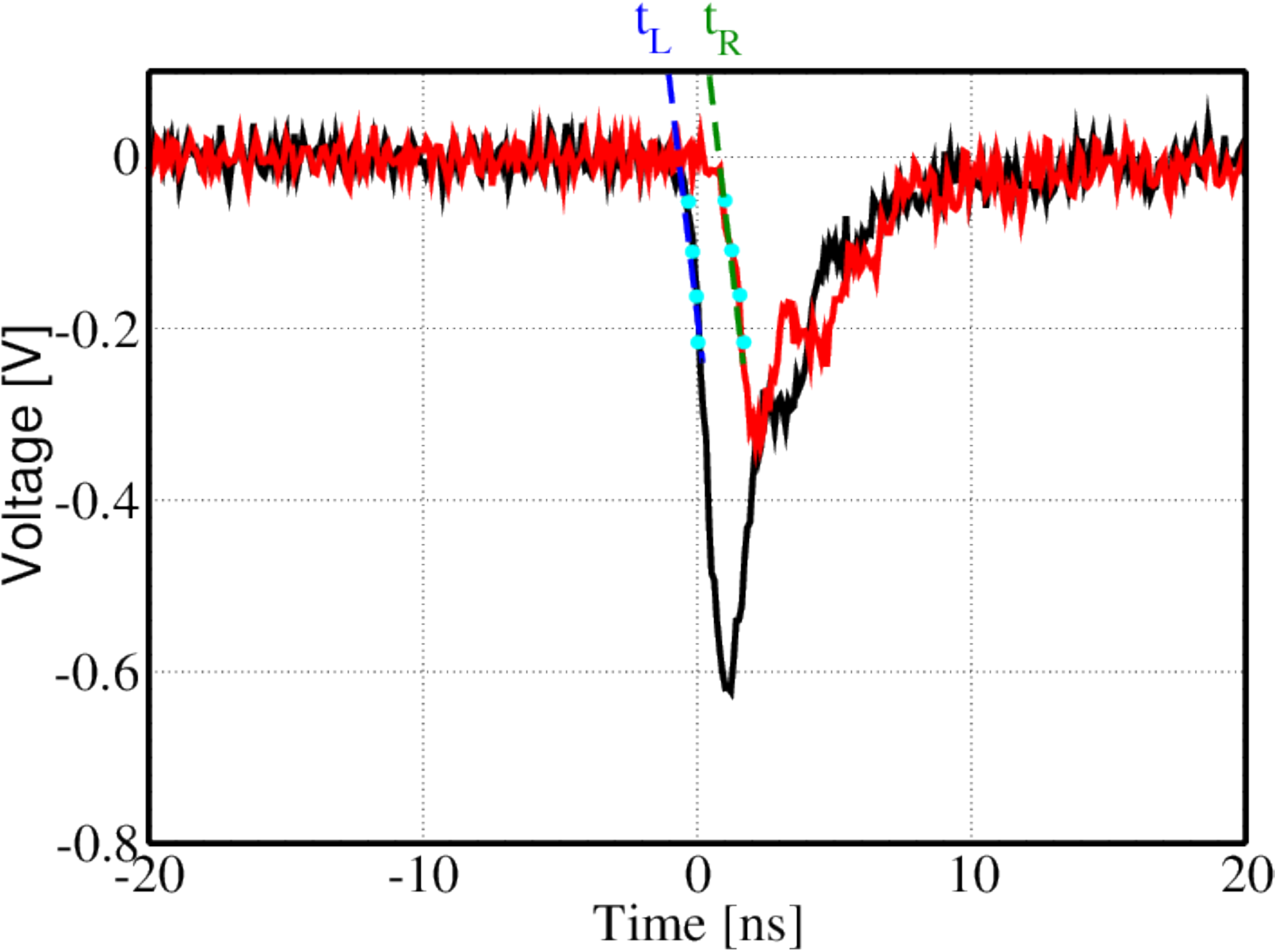}
    \caption{The Multi-Voltage-Threshold method applied to a sample event pulse. Black and red curves represent the  left and right signals of an event, respectively. Dots are the applied thresholds on both signals. The green and blue dashed lines are lines fitted to leading edges of the left and right signals passing through the applied threshold levels, respectively. The signal times (t$_L$ and t$_R$) are defined by zero value of the fitted lines and the event time is the average of t$_L$ and t$_R$ (left and right) times.}
    \label{fig:Fig10}
\end{figure}

In this article, in addition to method based on the Mahalanobis distance we also test another method for determining the time of the signals. The second method is based on a linear fit to the times measured at the leading edge analogously as applied in references~\cite{KIM2009B,XIE2009A,XIE2005}. The event time was then calculated as a zero value of a regression line fitted to the measured leading edge points. The scheme of the method is shown in Fig.\ref{fig:Fig10}.
 In Ref. \cite{XIE2009A}, this method was tested on the data obtained from a pair of LSO crystals of dimensions 6.25 x 6.25 x 25 mm$^3$. The scintillators were wrapped in the teflon tape and optically coupled to the photomultiplier tubes. Signals were readout via a Tektronix TDS6154C digital oscilloscope using coincident technique. A weak ${}^{18}$F source was used to irradiate the scintillators. The signals were sampled at number of pre-defined voltage levels and the best coincidence timing resolution of about 302 ps (FWHM) was obtained at 16 voltage threshold levels. 

In another work \cite{KIM2009B}, using ${}^{22}$Na source, again the signals were sampled at 50, 100, 200 and 300 mV threshold levels and the Multi-Voltage-Threshold was used to compute coincidence timing resolution. As a result 340 ps (FWHM) was obtained as coincidence timing resolution which is closer to the value obtained with digital library sampled at 20 GSps using multi-threshold discriminator board. Thus, this is also one of the potential methods for implementing digital PET data acquisition.

\section{Validation of the method}
\subsection{Data collection and filtration}
\begin{figure}[h]
  \centerline{\includegraphics[scale=0.24]{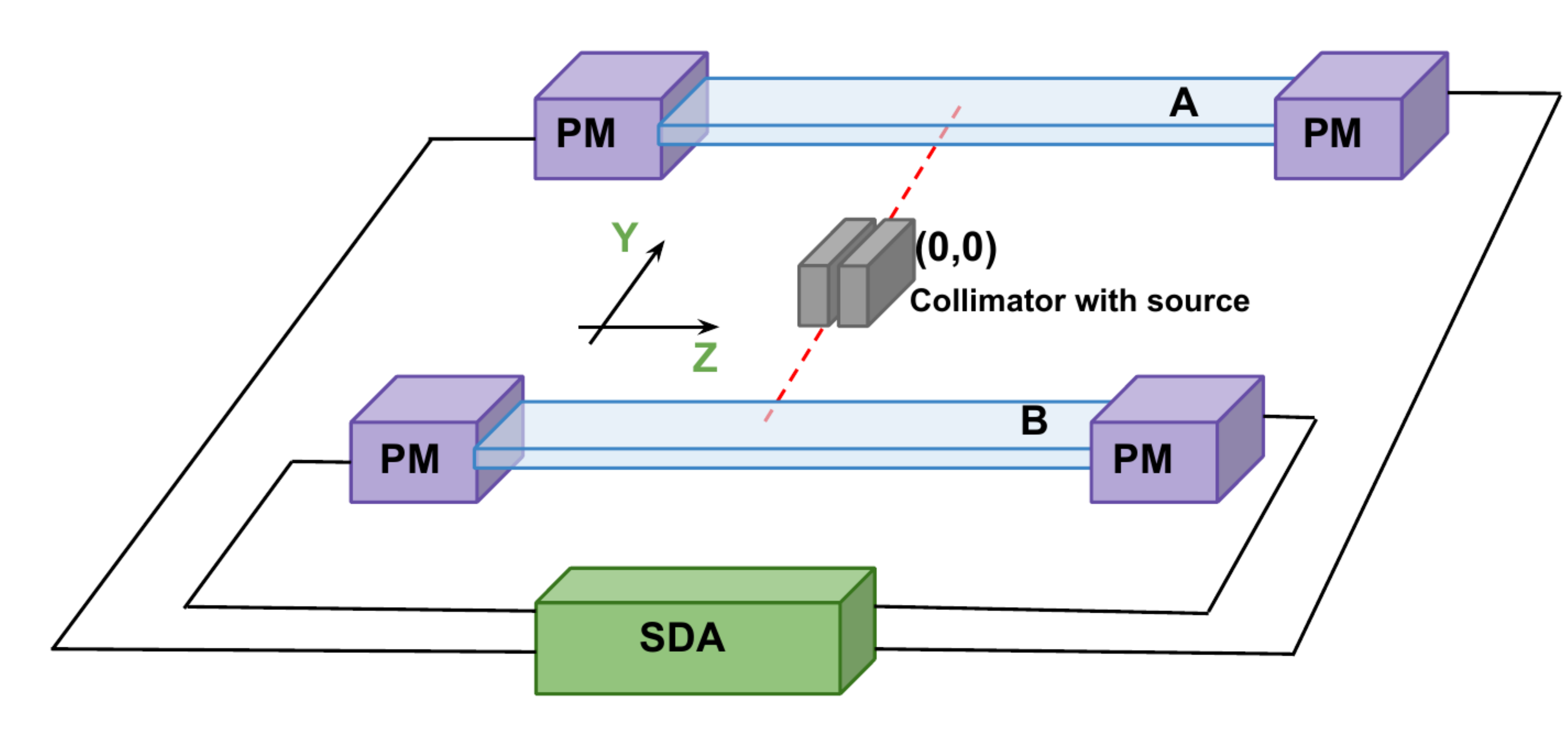}}
  \caption[]{A schematic view of the double-strip J-PET prototype used for the validation of the proposed method. The two detectors were red out by photomultipliers (PM) connected to the Serial Data Analyzer (Lecroy SDA6000A) providing coincident registration of all four signals. A collimated beam of annihilation gamma quanta was provided by a $^{22}$Na source placed at (0,0) position.}
  \label{fig:Fig11}
  \end{figure}

The method was validated on the same data measured with the double strip J-PET prototype (mentioned in sec.\ref{library}). The prototype was built out of two BC-420~\cite{SAINTGOBAIN} plastic scintillators with dimensions 300 x 19 x 5 mm$^3$. Both  strips were wrapped with 3M Vikuiti specular reflector foil~\cite{VIKUITI}. Signals from strips were read out using Hamamatsu photomultipliers R9800~\cite{HAMAMATSU} connected optically via optical grease EJ-550 to the ends of scintillators and probed by Serial Data Analyzer (Lecroy SDA6000A) with a time interval of 100~ps. The ${}^{22}$Na isotope was used as a source of annihilation photons. For noise suppression only coincident signals from both detectors were registered. The scheme of the used experimental setup is presented in Fig.\ref{fig:Fig11}. The measurements were done along the whole length of scintillator at positions for which we earlier determined the library of model signals. This allowed us to determine the achievable performance of the reconstruction in function of the hit position.  
%
%
After collection of the dataset a primary correction for pedestals was implemented on all the registered signals to have a signal library free of the electronic voltage offset. Next, only those events were considered for the reconstruction of hit-position and hit-time for which energy depositions were in the range from 200~keV to 380~keV. This is the range of energy loss which will be used for the J-PET tomography in order to minimise the blurring of the image due to the scatterings of gamma quanta in the patient~\cite{MOSKAL14C,KOWALSKI2018}. The relation between the measured charge and deposited energy was computed by fitting the Klein-Nishina formula~\cite{KLEIN-NISHA} convoluted with the detector resolution to the experimental energy loss spectrum~\cite{MOSKAL14C}. The data filtering was done to suppress most of those events which originate from secondary Compton scattering~\cite{KOWALSKI2016A} and also from the 1.27 MeV gamma quanta produced in the decay of ${}^{22}$Na isotope.
\subsection{Optimization of the signal processing}
The full-scale J-PET tomograph signals will be sampled by a dedicated front-end electronics (FEE) in voltage domain with a pre-defined set of thresholds with time resolution of about 20~ps~\cite{MAREK2014A}. In order to design the optimal configuration of thresholds for the full scale J-PET tomograph it is necessary to determine the optimal number of thresholds and voltage value for each threshold. To this end the measured data were optimized by constant level discriminator approach followed by energy deposition classifier using the Mahalanobis distance defined in Eq.~\ref{eq:Equation2}.
The Time-of-Flight resolution was used as a criterion for the optimization. We have chosen threshold levels at which root mean squared error value of TOF distribution i.e. rms(TOF) obtained from Mahalanobis distance is minimal~\cite{NehaPhdThesis}. We have performed tests with a different number of applied threshold levels but no significant improvement was observed on increasing the number of threshold levels by more than two. The reason may be the fact that the signal is composed of many single photoelectron pulses but only few of them contribute significantly to the onset of the leading edge~\cite{MOSKAL2016A}.
\begin{figure}[h]
  \centerline{\includegraphics[scale=0.36]{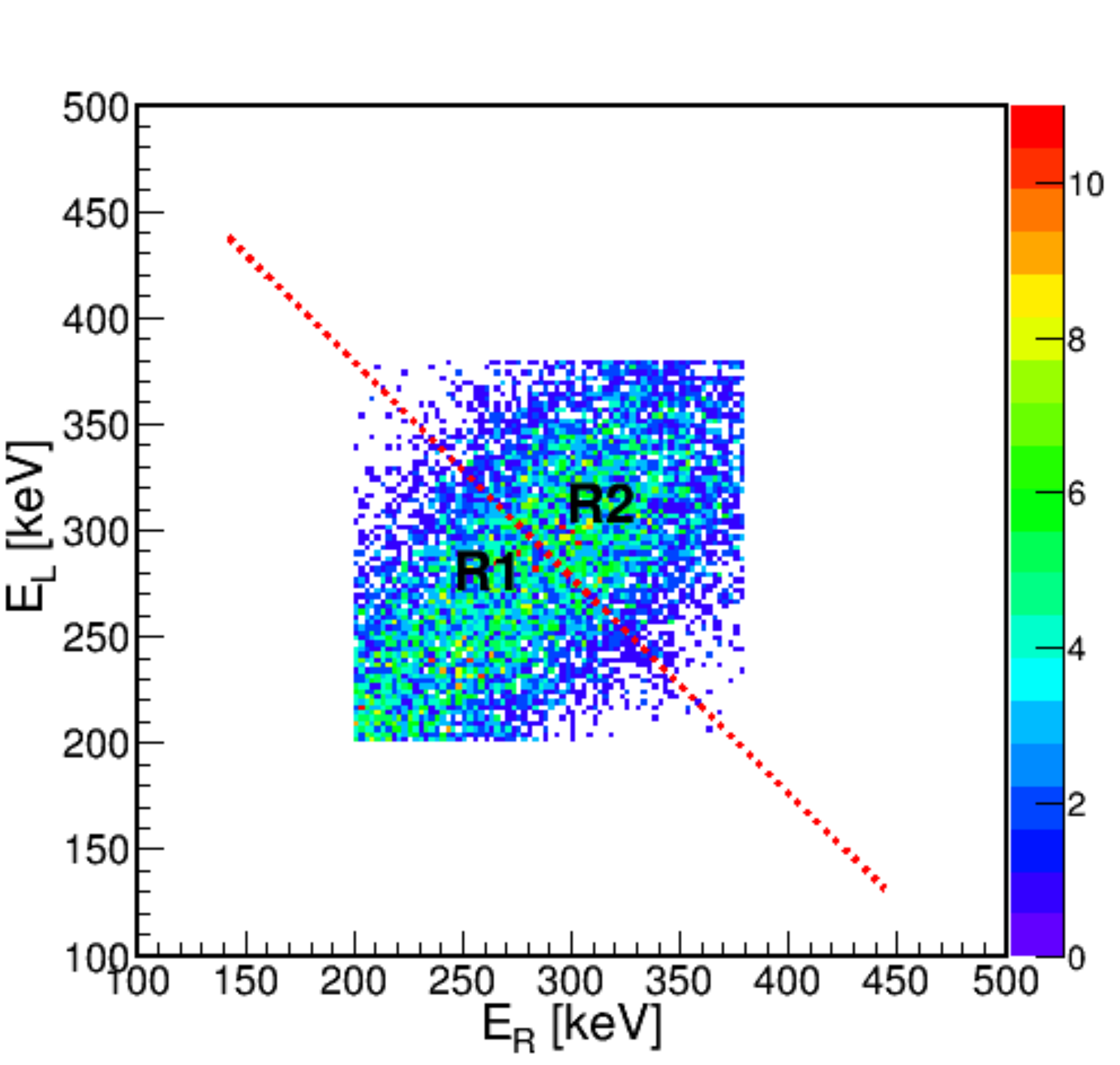}}
   \caption{Scatter plot of energy deposition registered at the left (E$_L$) and right (E$_R$) side of one of the J-PET detector modules. The red dotted line shows the bisection of energy loss into 2 parts (R1 and R2) done to improve the TOF resolution. Only energies in the range from 200~keV to 380~keV are shown.}
  \label{fig:Fig12}
\end{figure}
Thus, we have performed optimization of two-threshold levels for signals measured at the center of the scintillator. As a result, we obtained -55~mV and -100~mV as the optimized constant levels of discrimination and these values were used as the reference for all other hit-positions. It was noticed that the time resolution and, hence, the covariance matrix depends on the number of photoelectrons in the signal which corresponds to the energy deposited by the interacting gamma photon. Therefore, the comparison of signals should improve if the covariance matrix would be established as a function of the energy loss. To check this we divided the range of available energy losses (i.e. 200~keV to 380~keV) into several regions and computed the covariance matrix for each of them separetely. No significant improvement in the resolution was found after dividing the whole energy loss region to more than two parts~\cite{NehaPhdThesis}. Bisection of energy distribution into two parts (R1 and R2) is presented in Fig.\ref{fig:Fig12}.
\section{Results}
\subsection{Mahalanobis method}
\begin{figure}[h]
    \centering{\includegraphics[scale=0.38]{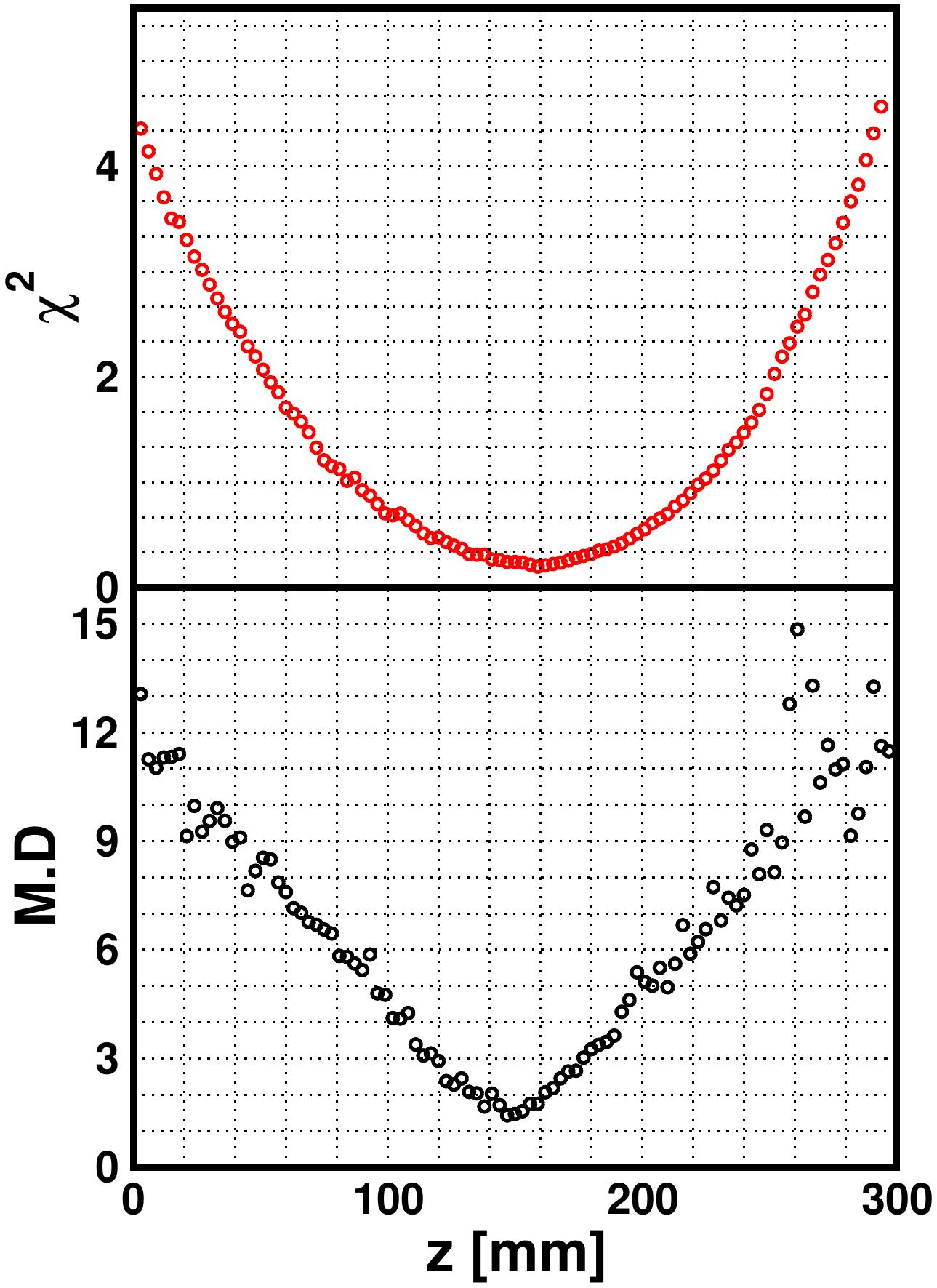}}
     \caption{Position of an event that belongs to the central hit-position i.e. 150~mm reconstructed by two methods: (upper panel) $\chi^2$ minimization~\cite{NEHA2015A}  and (lower panel) Mahalanobis distance. }
     \label{fig:chi-sq_MD_comp}
\end{figure}

The Mahalanobis distance distribution calculated according to Eq.\ref{eq:Equation2} for an event from the central hit-position sampled at pre-defined two-threshold levels is shown in lower panel of Fig.\ref{fig:chi-sq_MD_comp}. The upper panel of Fig.\ref{fig:chi-sq_MD_comp} is showing the $\chi^2$ distribution of the same event presented in~\cite{NEHA2015A}. One can see clear the value of position reconstructed by Mahalanobis distance has less uncertainty then the one reconstructed by $\chi^2$ minimization method.

\begin{figure}[h]
\centering
  \subfloat[\label{Fig13a}]{\includegraphics[width=4.19cm,height=4.15cm]{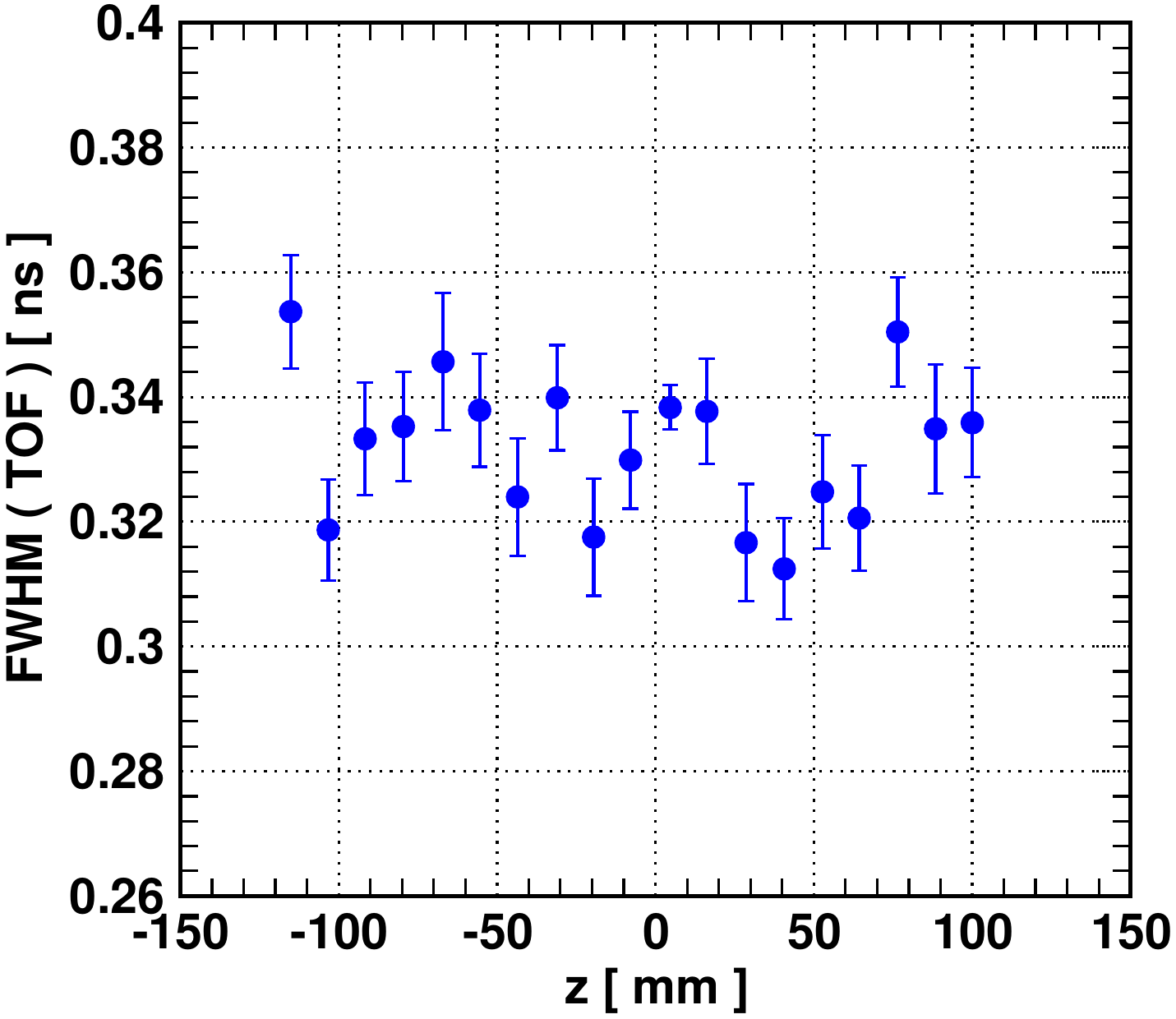}}
  \hfill
  \subfloat[\label{Fig13b}]{\includegraphics[width=4.14cm,height=4.15cm]{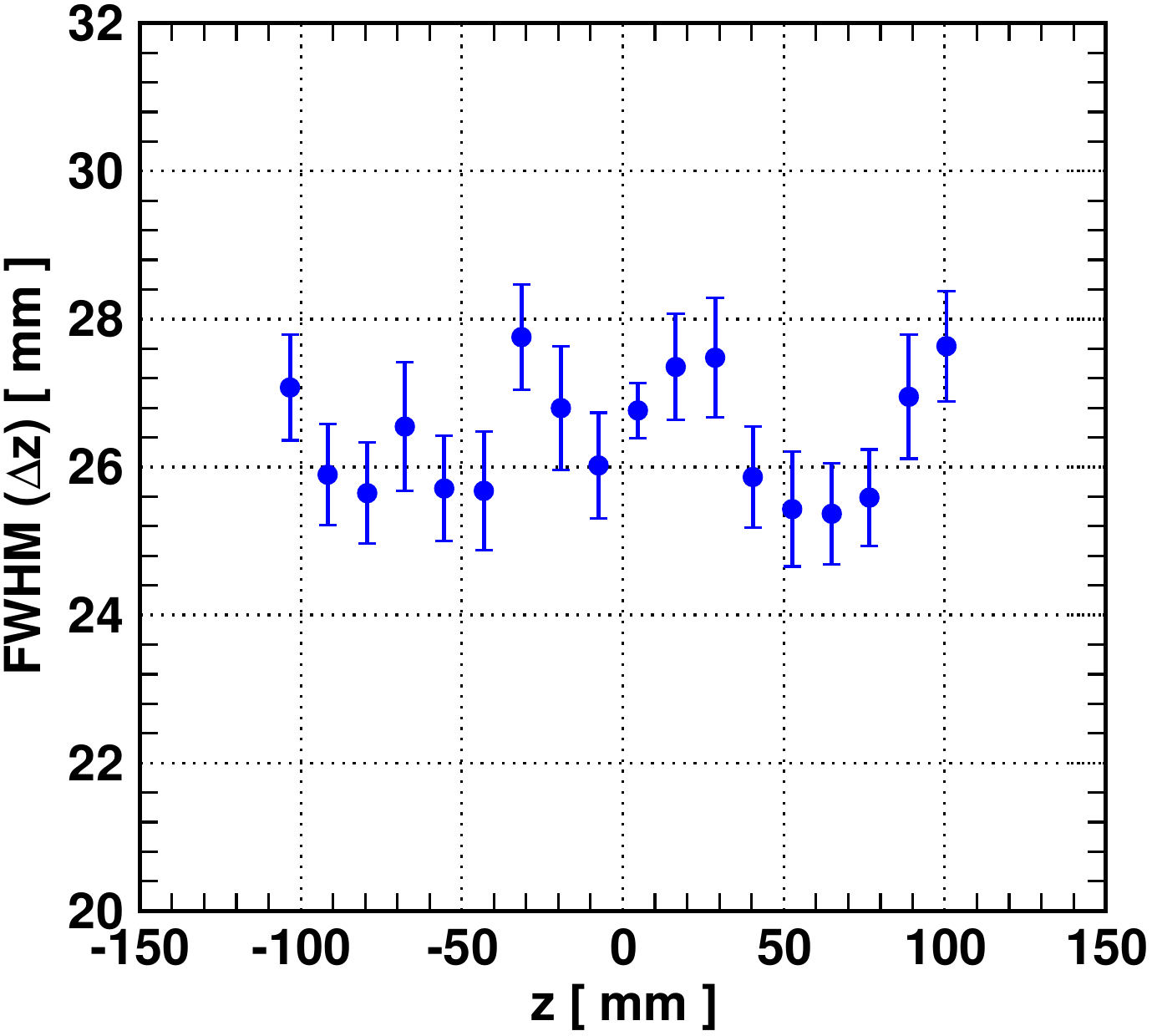}}
    \caption{(a) TOF resolution as a function of the hit-position (z) along the detector. (b) Spatial ($\Delta$z) resolution as a function of the hit-position (z) along the detector.}
    \label{fig:Fig13}
\end{figure}

\begin{figure}[h]
  \centering
  \subfloat[\label{Fig14a}]{\includegraphics[width=4.19cm,height=4.15cm]{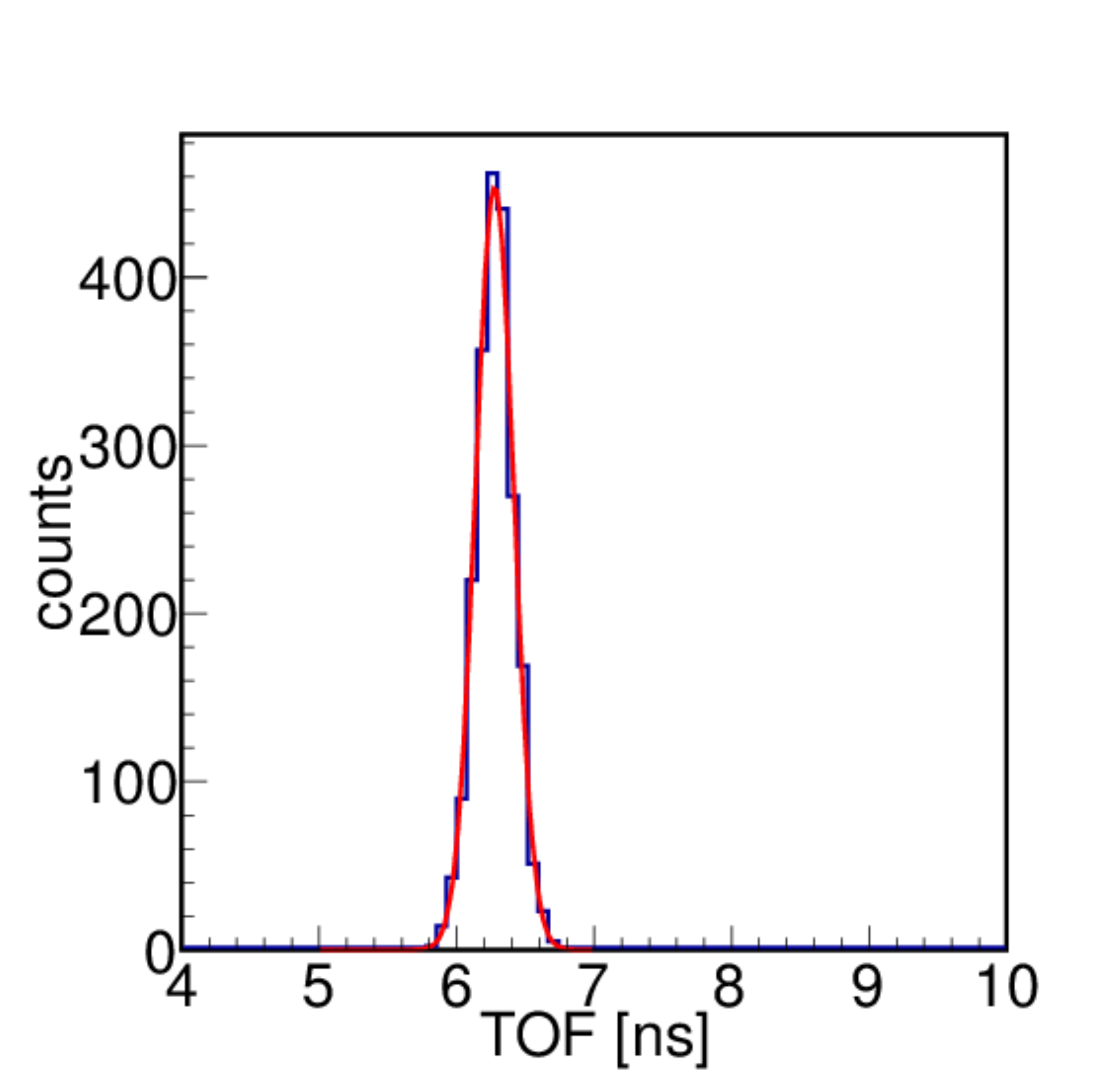}}
  \hfill
  \subfloat[\label{Fig14b}]{\includegraphics[width=4.14cm,height=4.1cm]{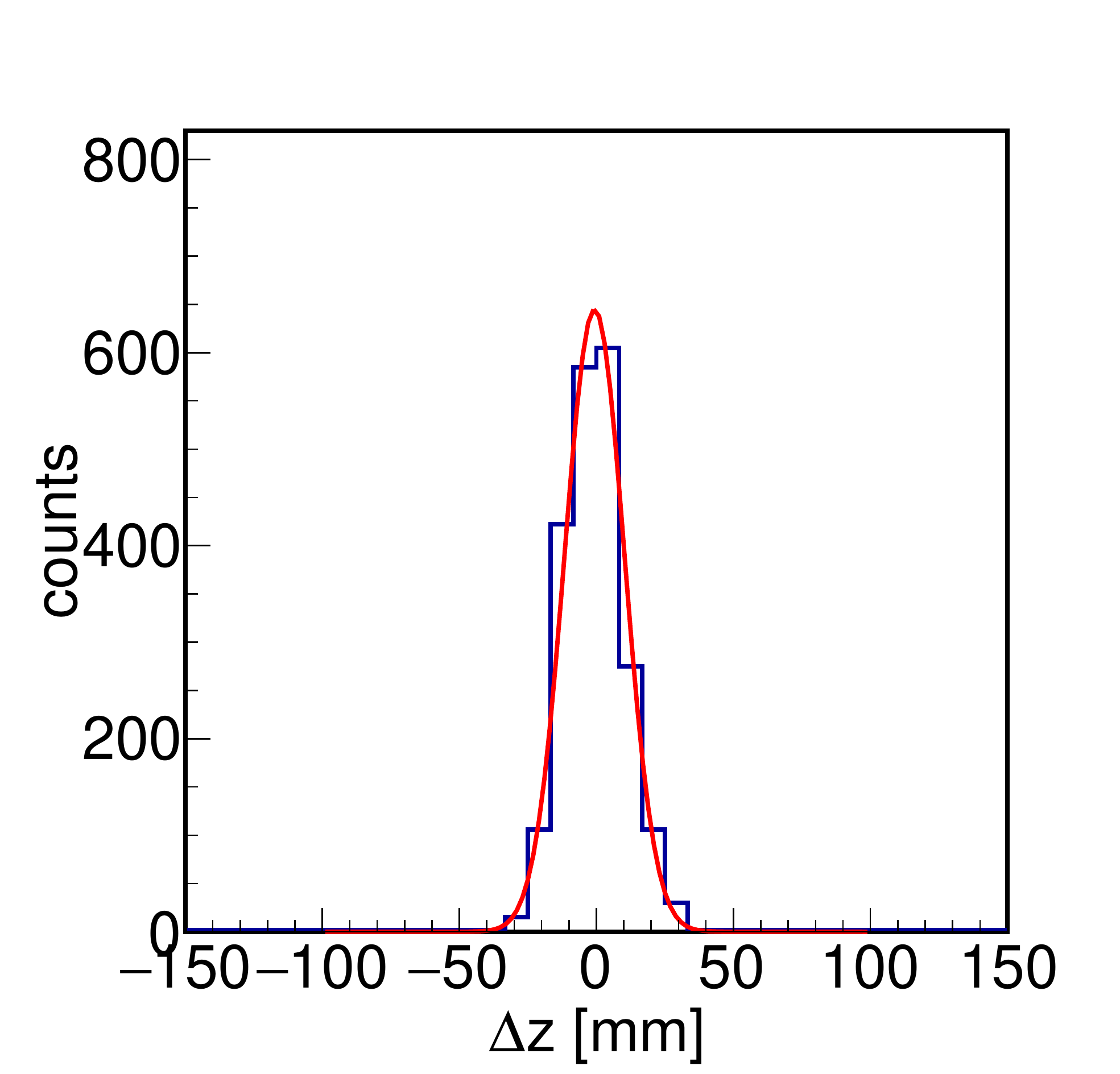}}
  \caption{(a) TOF distribution at the central hit-position with rms~(TOF)~= 0.138~$\pm$~0.002~ns. The mean of TOF distribution is not zero because of offsets due to different length of cables used to readout the photomultipliers output. (b) Spatial ($\Delta$z) distributions at central hit-position with rms~($\Delta$z)~=~10.7~$\pm$~0.2~mm. }
  \label{fig:Fig14}
\end{figure}

The Time-of-Flight and spatial resolutions obtained for hit-positions in the range from (-100, 0) mm to (100, 0) mm along the length of scintillator using the Mahalanobis method are presented in Fig.\ref{Fig13a} and Fig.\ref{Fig13b}, respectively. Exemplary Time-of-Flight and spatial ($\Delta$z) distributions at optimized levels followed by energy classifier are shown in Fig.\ref{Fig14a} and Fig.\ref{Fig14b}, respectively. It is evident from Fig.\ref{Fig13a} and Fig.\ref{Fig13b} that both resolutions remain constant at all z values  along the whole length of the scintillator strip. The achieved Time-of-Flight resolution along 300 mm long scintillator amounts to 325~ps~(FWHM) and the spatial resolution is 25~mm~(FWHM).

\begin{figure}[h]
    \centering{\includegraphics[scale=0.20]{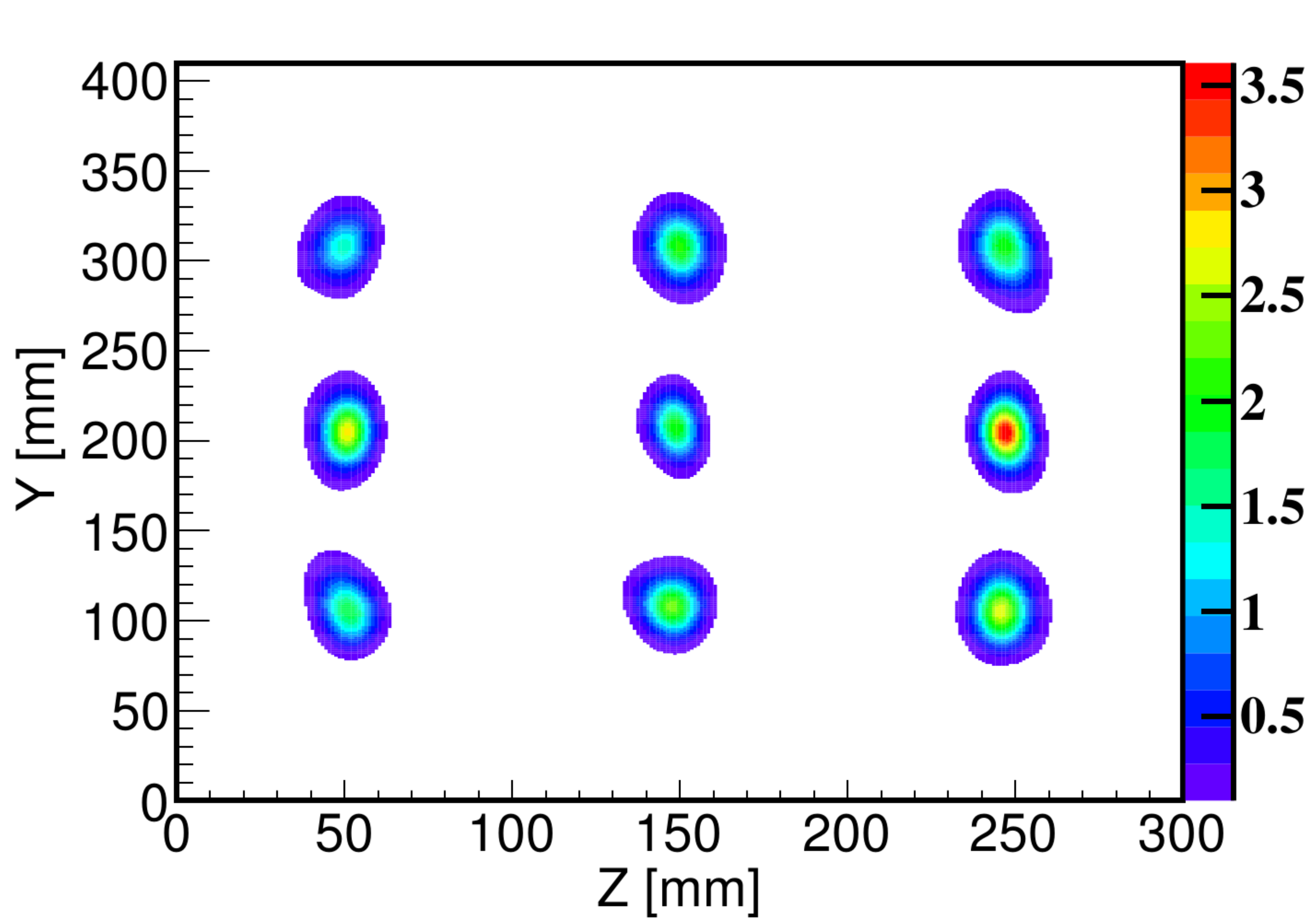}}
     \caption{2D image reconstructed for the sources placed at nine different positions using the MLEM algorithm. Each point represents sources of different activity placed in between the two strips. The source at position (250, 200)~mm was the most active one.}
     \label{fig:Fig15}
\end{figure}
Moreover, in order to test the method in view of its application in Positron Emission Tomography, we performed experiments with bare $^{22}$Na sources emitting annihilation gamma quanta isotropically. The activities of the used sources are listed in Table.\ref{tab:table1}. The sources were placed at nine different positions between the two scintillators. The scheme of source position is presented in~\cite{NehaPhdThesis}. The reconstructed image of these sources using maximum likelihood expectation maximization (MLEM) algorithm~\cite{SLOMSKI2014,BIALAS2015A,ADAMTHESIS} is presented in Fig.\ref{fig:Fig15}. Even though only two strips were used to detect the signals it was possible to reconstruct a tomographic image. The point spread functions in transverse and axial directions, calculated according to the NEMA norm~\cite{NEMA}, are equal to FWHM~=~20.2~mm and FWHM~=~7.7~mm, respectively.\\
Indeed, the axial and transaxial resolutions obtained for the two-strip J-PET prototype are worse in comparison to commercial tomographs. This was expected, since we have used only one pair of scintillators. However, we expect the further improvement when the image will be done with scintillators forming the cylindrical barrel providing LOR, and TOF in three dimensions~\cite{KOWALSKI2018}.

\begin{table}
    \caption{List of different ${}^{22}$Na sources used for the test measurement.}
    \label{tab:table1}
    \begin{center}
    \begin{tabular}{ |c|c| }
    \hline
     Source No. & Activity [kBq] \\ 
     \hline
     1 & 393  \\  
     2 & 140   \\
     3 & 399  \\
     4 & 391  \\
     5 & 2180  \\
     6 & 11946 \\
     7 & 2185  \\
     8 & 280  \\
     9 & 9  \\
     \hline
    \end{tabular}
    \end{center}
\end{table}

\subsection{Comparison of the Mahalanobis reconstruction with Linear Fitting method}
\begin{figure}[h]
  \centerline{\includegraphics[scale=0.40]{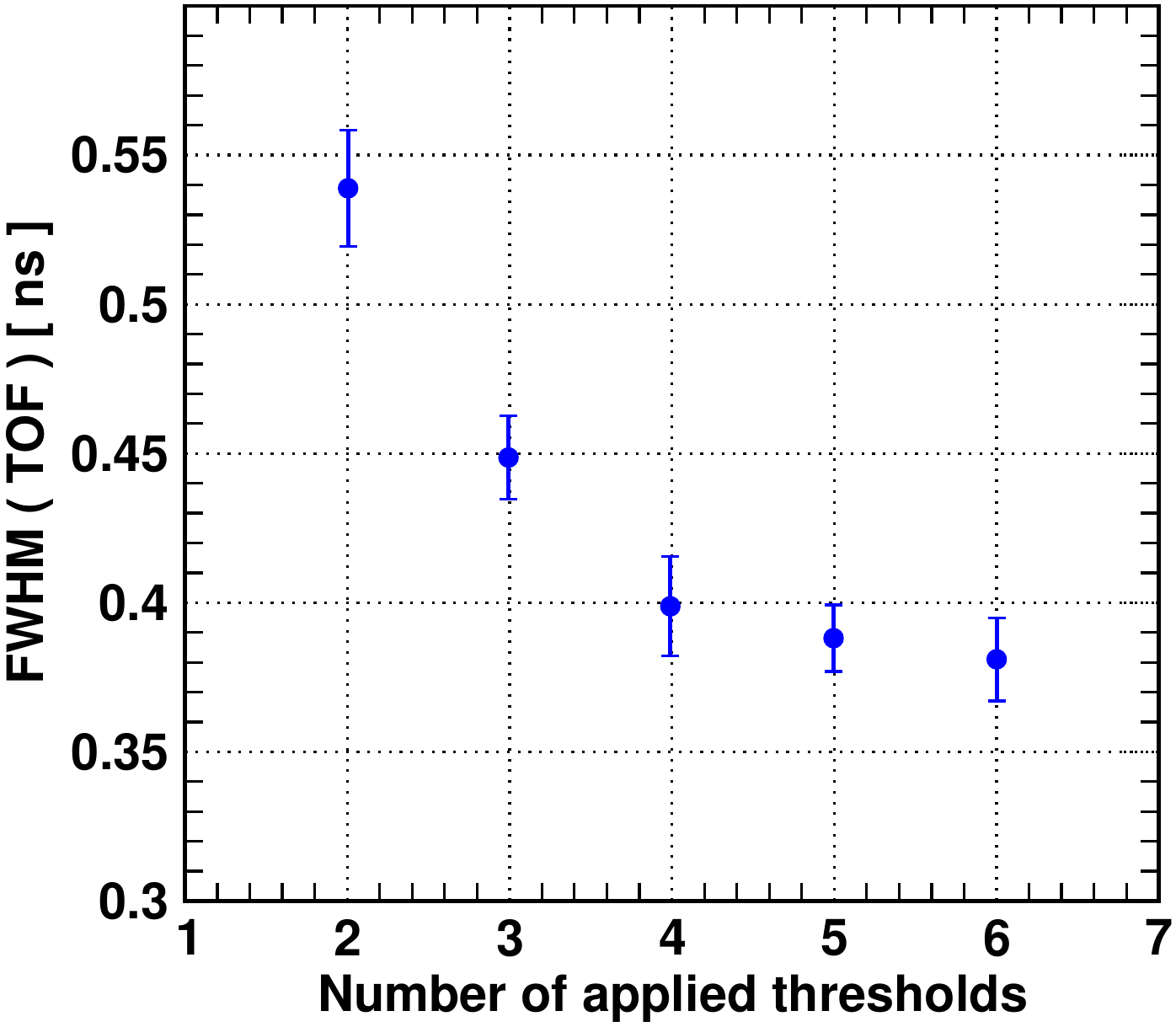}}
  \caption[]{Time-of-Flight resolution as a function of the number of applied threshold levels obtained using the Linear Fitting Method. The values of applied thresholds for each set is presented in Tab.~\ref{tab:MVT_Levels_combinations}.}
  \label{fig:Fig16}%
  \end{figure}

In order to test the Linear Fitting method we have determined the optimal number of applied thresholds looking for the best achievable Time-of-Flight resolution. The test was performed with different number of threshold levels. For each number of levels different values of threshold were tested e.g. for two-threshold levels the value of TOF was calculated at a number of different set of values: (-40 mV, -80 mV), (-40 mV, 100 mV), (-60 mV, -100 mV), (-80 mV, -160 mV) etc. In Tab.~\ref{tab:MVT_Levels_combinations} only those values for each number of applied levels are given for which we obtained best TOF resolution. The dependence of TOF resolution on the number of applied threshold levels is shown in Fig.\ref{fig:Fig16}. One can see that the resolution improves with the number of applied thresholds, but after discrimination on four of them there is no significant improvement with the current threshold setting in our experiment. The obtained value of rms~(TOF) with four pre-defined threshold levels is comparable with the value calculated using the Mahalanobis method at two pre-defined threshold levels.

\begin{table}[h!]
\centering
  \caption{Number of applied threshold levels with their values for which the best TOF resolution was obtained.}
 \label{tab:MVT_Levels_combinations}
 \vspace*{0.5cm}
\begin{tabular}{|c||c|}
\hline\hline 
Number of applied levels  & Values [mV] \\[0.5ex]
\hline 
2 & -40, -80   \\ 
3 & -40, -80 , -120      \\
4 & -40, -80 , -120, -160   \\
5 & -40, -80 , -120, -160, -200   \\
6 & -40, -80 , -120, -160, -200, -240   \\
7 & -40, -80 , -120. -160, -200, -240, 280   \\
\hline
\end{tabular}
\end{table}
Although the algorithm of Linear Fitting method is much easier to apply than the algorithm of the Mahalanobis method, its realization is impractical because it requires four thresholds to achieve the same resolution as with two thresholds when applying the method based on the Mahalanobis distance. A higher number of applied threshold levels would increase the cost of electronics used in the PET scanner and would make it more expensive. 

\section{Summary}
\begin{figure}[h]
 \centering
  \subfloat[\label{Fig17a}]{\includegraphics[width=4.20cm,height=4.1cm]{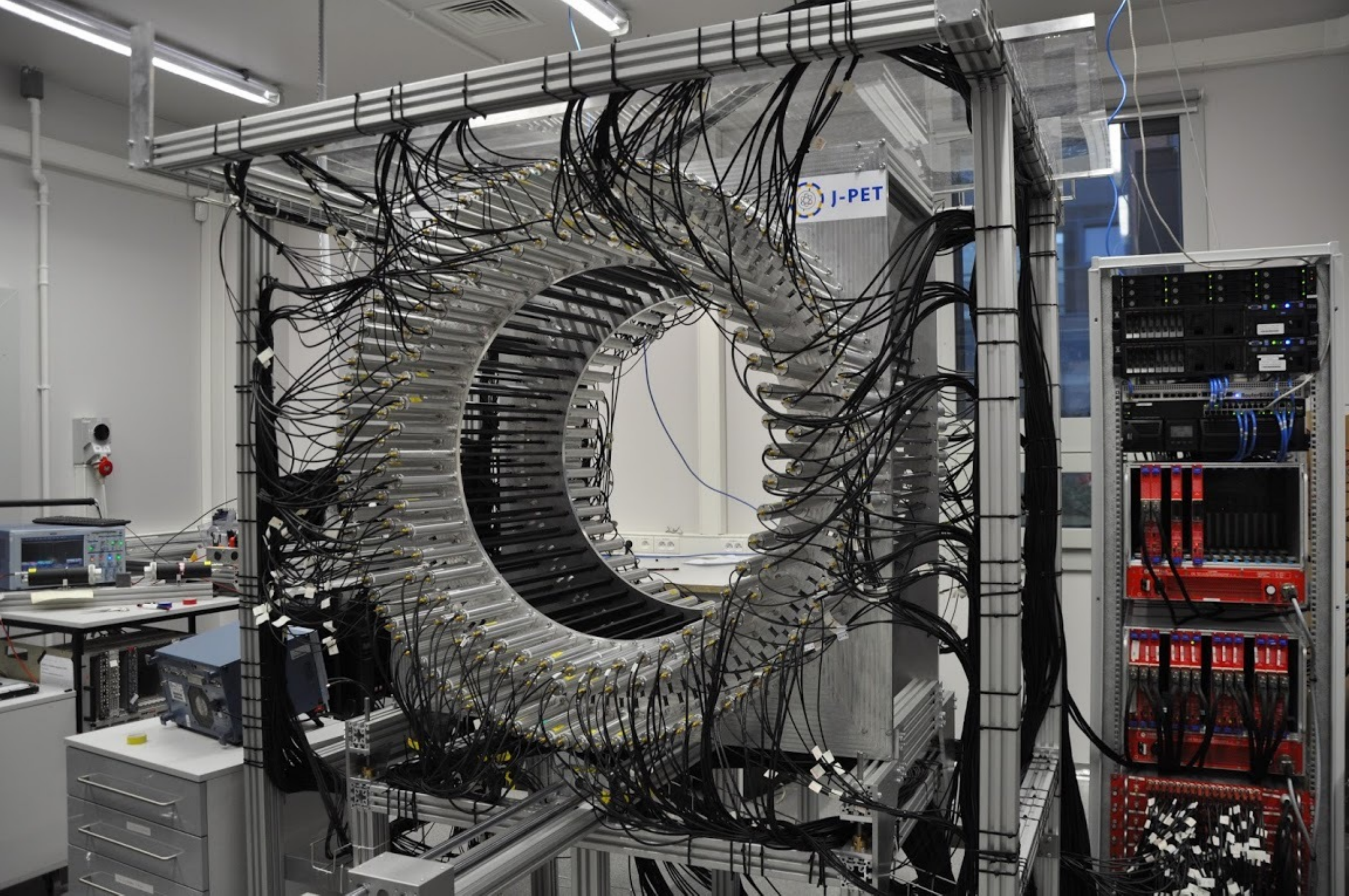}}
  \hfill
  \subfloat[\label{Fig17b}]{\includegraphics[width=4.20cm, height=4.1cm]{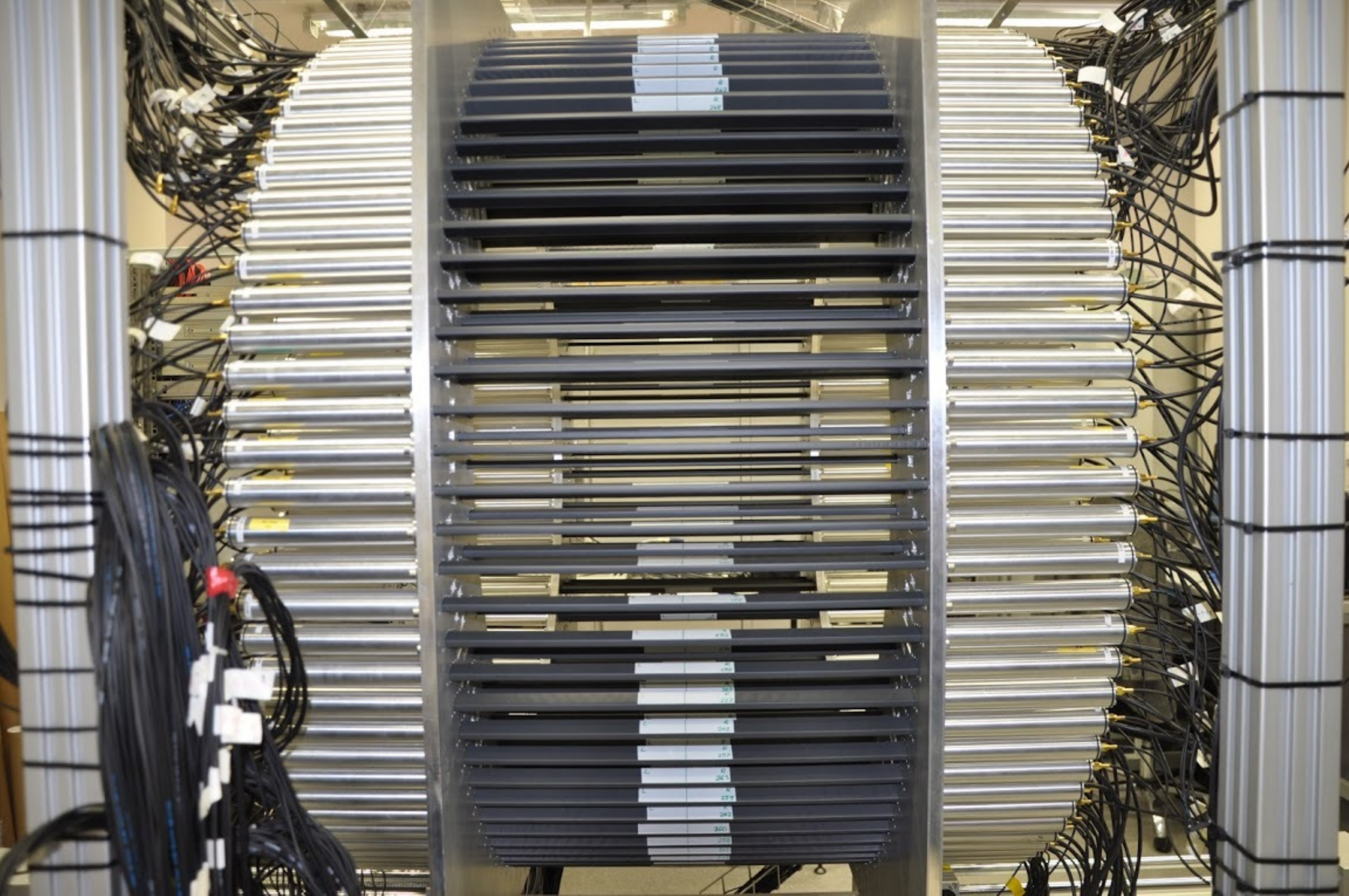}}
  \caption{(a) Front view of the full frame J-PET prototype with large field of view (diameter of 85~cm) and axial length of 50~cm. (b) Rear view of the full frame J-PET prototype with large field of view (diameter of 85~cm) and axial length of 50~cm. }
  \label{fig:Fig17}
\end{figure}

An optimized method of reconstructing the hit-position and hit-time of photons in scintillator detectors  was developed in view of its application for the registration of 511~keV photons in the Positron Emission Tomography. The method was validated on experimental data collected with the double-strip J-PET prototype. It is based on the comparison of the measured signal with synchronized model signals stored in a library utilizing the Mahalanobis distance as a measure of similarity. The reconstructed hit-position of the photons is considered as the position of a library signal most similar to the measured one, and the hit-time is the relative time difference between them. The time difference between the reconstructed hit-time (interaction time) of photons in different detectors corresponds to the Time-of-Flight (TOF) which is independent of the trigger time. The best obtained Time-of-Flight and spatial resolutions amount to 325~ps (FWHM) and 25~mm~(FWHM), respectively. This result is comparable to the one obtained from Linear Fitting method requiring higher number of pre-defined thresholds than two.\\ 
In Summary, the cost of the electronics for a J-PET scanner device depends on the number of voltage thresholds levels needed in the reconstruction of the time and spatial resolution. In this contribution we have studied two methods and compared them with respect to the obtained resolution. We found that the Mahalanobis method needs less levels than the Linear Fitting method in obtaining the same resolution. In a next step we will apply the Mahalanobis method for all 192 scintillators which build up the J-PET full frame detector positioned along a cylindrical geometry in 3 layers~\cite{SZYMON2017}. The front and rear view of the J-PET full frame prototype are shown in Fig.\ref{Fig17a} and Fig.\ref{Fig17b}, respectively. 

\section*{Acknowledgement}

The authors acknowledge technical and administrative support of A. Heczko, M. Kajetanowicz and W. Migda\l{}. This work was supported by The Polish National Center for Research
and Development through grant INNOTECH-K1/IN1/64/159174/NCBR/12, the
Foundation for Polish Science through the MPD and TEAM POIR.04.04.00-00-4204/17 programmes, the National Science Centre of Poland through grants no.\
2016/21/B/ST2/01222,\linebreak[3] 2017/25/N/NZ1/00861,
the Ministry for Science and Higher Education through grants no. 6673/IA/SP/2016,
7150/E-338/SPUB/2017/1 and 7150/E-338/M/2017, the Austrian Science Fund FWF-P26783 and DSC grants. We are also grateful to Prof. Steven Bass for reading and correcting the manuscript.


\bibliographystyle{IEEEtran}
\bibliography{IEEE_ref}

\end{document}